\documentclass[iop]{emulateapj}
\usepackage{amssymb}
\usepackage{amsmath}
\usepackage{rotating}
\usepackage{longtable}

\newcommand{\msun}{M_\odot}
\newcommand{\lsun}{L_\odot}

\begin{document}
\title{Variable Stars in Large Magellanic Cloud Globular Clusters II: NGC 1786*}
\author{Charles A. Kuehn\altaffilmark{1,2}, Horace A. Smith\altaffilmark{1}, M\'arcio Catelan\altaffilmark{3,4}, Barton J. Pritzl\altaffilmark{5}, Nathan De Lee\altaffilmark{1,6}, Jura Borissova\altaffilmark{7}}
\altaffiltext{*}{Based on observations taken with the SMARTS $1.3$-meter telescope operated by the SMARTS Consortium and observations taken at the Southern Astrophysical Research (SOAR) telescope, which is a joint project of the Minist\'{e}rio da Ci\^{e}ncia, Tecnologia, e Inova\c{c}\~{a}o (MCTI) da Rep\'{u}blica Federativa do Brasil, the U.S. National Optical Astronomy Observatory (NOAO), the University of North Carolina at Chapel Hill (UNC), and Michigan State University (MSU).}\altaffiltext{1}{Department of Physics and Astronomy, Michigan State University, East Lansing, MI 48824, USA; smith@pa.msu.edu}
\altaffiltext{2}{Current address: Sydney Institute for Astronomy, University of Sydney, Sydney, Australia; kuehn@physics.usyd.edu.au}
\altaffiltext{3}{Pontificia Universidad Cat$\rm{\acute{o}}$lica de Chile, Facultad de F\'{i}sica, Departamento de Astronom\'ia y Astrof\'isica, Santiago, Chile; mcatelan@astro.puc.cl}
\altaffiltext{4}{The Milky Way Millennium Nucleus, Santiago, Chile}
\altaffiltext{5}{Department of Physics and Astronomy, University of Wisconsin Oskosh, Oshkosh, WI 54901, USA; pritzlb@uwosh.edu}
\altaffiltext{6}{Current address: Department of Physics and Astronomy, Vanderbilt University, Nashville, TN 37235, USA; nathan.delee@vanderbilt.edu}
\altaffiltext{7}{Departamento de F\'isica y Astronom\'ia, Falcultad de Ciencias, Universidad de Valpara\'iso, Valpara\'iso, Chile; jura.borissova@uv.cl}

\begin{abstract}
This is the second in a series of papers studying the variable stars in Large Magellanic Cloud globular clusters.  The primary goal of this series is to study how RR Lyrae stars in Oosterhoff-intermediate systems compare to their counterparts in Oosterhoff I/II systems.  In this paper, we present the results of our new time-series BV photometeric study of the globular cluster NGC 1786.  A total of 65 variable stars were identified in our field of view.  These variables include $53$ RR Lyraes ($27$ RRab, $18$ RRc, and $8$ RRd), $3$ classical Cepheids, $1$ Type II Cepheid, $1$ Anomalous Cepheid, $2$ eclipsing binaries, $3$ Delta Scuti/SX Phoenicis variables, and $2$ variables of undetermined type.  Photometeric parameters for these variables are presented.  We present physical properties for some of the RR Lyrae stars, derived from Fourier analysis of their light curves.  We discuss several different indicators of Oosterhoff type which indicate that the Oosterhoff classification of NGC 1786 is not as clear cut as what is scene in most globular clusters.
\end{abstract}


\section{Introduction}

This is the second in a series of papers focusing on the variable stars in globular clusters in the Large Magellanic Cloud (LMC).  The goal of this series is to better understand the Oosterhoff dichotomy, the tendency for Milky Way globular clusters to be classified as either Oosterhoff I (Oo-I) or Oosterhoff II (Oo-II) objects, and how the Oosterhoff intermediate clusters in the nearby dwarf galaxies fit into that picture.  The Oosterhoff groups are traditionally defined based on the average periods of the RRab stars with Oo-I clusters having an average RRab period of $\langle P_{ab}\rangle < 0.58$ days and Oo-II clusters having $\langle P_{ab}\rangle >0.62$ days.  In general Oo-II clusters also tend to be more metal-poor and have a larger ratio of first overtone to fundametal mode RR Lyrae.  The period range between the two groups, $0.58\leq\langle P_{ab}\rangle \leq 0.62$ days, is referred to as the Oosterhoff gap and is essentially unoccupied by Milky Way globular clusters.  The nearby dwarf galaxies and their globular clusters present a sharp contrast to this behavior as these extra-galactic clusters not only fall in the Oo-I and Oo-II groups, they also fall into the gap between these groups; in fact the extra-galactic objects seem to preferentially be located in the gap.  These Oosterhoff-intermediate objects, as objects that fall in the gap are called, present a challeng for models that propose that the Milky Way halo was formed through the accretion of objects similar to the present day nearby dwarf galaxies as we would expect to see similar Oosterhoff properties in both samples if that were the case \citep{mc09}.  The present understanding of the Oosterhoff phenomenon will be discussed in more detail in a future paper in this series.  The first paper in this series, \citet{kuehn11}, discussed the variable stars in the LMC globular cluster NGC 1466.  In this paper we identify, classify, and discuss the variable stars found in the LMC globular cluster NGC 1786.

NGC 1786 is an old globular cluster that is located near the center of the LMC.  \citet{mucciarelli09} found a metallicity of ${\rm [Fe/H]}=-1.75\pm0.01$ from high-resolution spectroscopy of seven red giant stars.  However, \citet{olszewski91} found a slightly lower value of $-1.87\pm0.20$ from the low-resolution spectra of two giant stars; these values are consistent within the error bars.  \citet{sharma10} reported that NGC 1786 is not very reddened, having an E(B-V) = $0.06$, and has an age of $12.3$ Gyr.  This age is consistent with \citet{johnson02} finding, through the comparison of the cluster color-magnitude diagram (CMD) with theoretical isochrones, that NGC 1786 formed within $2$ Gyr of the other old LMC clusters .

NGC 1786 is a fairly centrally concentrated cluster which, combined with its location near the center of the LMC, makes it difficult to get accurate photometry of the stars in the center due to blending.  \citet{graham85} was the first to detect variable stars in NGC 1786, finding $12$ variables, for which he determined periods of $10$.  \citet{walker88} later observed NGC 1786 and confirmed the variable nature of $11$ of Graham's stars, though they were also only able to determine periods for ten of the stars.  Of the variables with known periods, $4$ were classified as RRab stars, $5$ as RRc stars, and one as a likely Anomalous Cepheid (AC).

More recently the Optical Gravitational Lensing Experiment-III (OGLE-III) observed the central portion of the LMC and many of its associated globular clusters, including NGC 1786 \citep{udalski08}.  The OGLE-III survey is designed to find instances of gravitational lensing, but a natural consequence of this project is the discovery and monitoring of variable stars.  OGLE-III greatly increased the number of identified variable stars in NGC 1786.  A total of $55$ RR Lyrae stars within $2$ arcsminutes of the cluster center were reported by \citet{soszy09}: $28$ RRab, $18$ RRc, and $9$ RRd stars.  The location of NGC 1786 near the center of the LMC means that some of these RR Lyrae stars likely belong to the field population of the LMC and not to the cluster.

%It should be noted that the cluster radius of $2$ arcminutes used by \citet{soszy09} was determined by visual inspection of images by \citet{bica08} and is not the same as the tidal radius of $2.9$ arcminutes determined by \citet{mclaughlin05} which is used in this paper.

In this paper we present the results of a new, ground-based search for variable stars in NGC 1786.  We use the RR Lyrae in the cluster to determine the distance modulus and Oosterhoff classification of the cluster.  We also present physical properties for the RR Lyrae stars determined through Fourier fitting of their light curves.

\section{Observations and Data Analysis}

A total of $57$ $V$ and $56$ $B$ images were obtained using the SOI imager ($5.2$x$5.2$ arcminute field of view) on the SOAR $4$-m telescope in November of 2007 and February of 2008 while $43$ $V$ and $42$ $B$ images were obtained using ANDICAM ($6$x$6$ arcminute field of view) on the SMARTS $1.3$-m telescope operated by the SMARTS consortium\footnote[8]{www.astro.yale.edu/smarts} from September 2006 to January 2007.   SOAR exposure times were between $30$s and $600$s for $V$ and between $60$s and $900$s for $B$.  SMARTS exposure times were $450$s in each filter.  Table \ref{obslog} lists the UT dates and times and time of each observation, which telescope it was on, the filter, and the seeing.

\begin{deluxetable*}{lccccc}
\tablewidth{0pc}
\tabletypesize{\scriptsize}
\tablecaption{Observing Log}
\tablehead{\colhead{DATE (UT)} & \colhead{TIME (UT)} & \colhead{FILTER} & \colhead{Exp Time (s)} & \colhead{Seeing (arcseconds)} &\colhead{Telescope}}

\startdata
2006-12-16& 01:38:35& B& 450.0& 2.16& SMARTS\\
2006-12-16& 01:47:03& V& 450.0& 1.83& SMARTS\\
2006-12-22& 01:24:48& B& 450.0& 1.54& SMARTS\\
2006-12-22& 01:33:16& V& 450.0& 1.12& SMARTS\\
2006-11-18& 02:51:41& B& 450.0& 2.01& SMARTS\\
\enddata
\tablecomments{This table is published in its entirety in the electronic edition.}
\label{obslog}
\end{deluxetable*}

Data reduction and variable identification were carried out in the manner described in the first paper in the series, \citet{kuehn11}.  Photometry was converted to the standard system using standard stars from \citet{stetson00} which were located within the field of NGC 1786. Of the $25$ Stetson standard stars that appeared in our observations, the average difference between our photometry and that of Stetson is $0.002\pm0.008$ mags in $V$ and $0.002\pm0.011$ mags in $B$. The majority of the identified variable stars were found in both the SOAR and SMARTS data however there was some disagreement between the two data sets existed for some of the more crowded stars.  In the cases where there was disagreement we used the SOAR results due to its better resolution.  The extremely crowded nature of NGC 1786 meant that the image subtraction package ISIS \citep{alard00} had to be used extensively to locate variable stars in the center of the cluster.  As stated in \citet{kuehn11}, periods of the variable stars are typically good to $\pm 0.0001$ or $0.0002$ days.  Aliasing with the RRd stars limits derived period ratios to an accuracy of $0.001$.

\section{Variable  Stars}

\begin{deluxetable*}{lcccccccccc}
\tablewidth{0pc}
\tabletypesize{\scriptsize}
\tablecaption{Photometric Parameters for Variables in NGC 1786, Excluding RRd Stars}
\tablehead{\colhead{ID}& \colhead{RA (J2000)} & \colhead{DEC (J2000)} & \colhead{Type}& \colhead{$P$ (days)}& \colhead{$A_{V}$}& \colhead{$A_{B}$}& \colhead{$\langle V\rangle$}& \colhead{$\langle B\rangle$}& \colhead{$\langle B-V\rangle$} & \colhead{OGLE-III ID}}

\startdata
V01& 	04:58:51.3&  $-$67:44:23.7&   AC&       	0.80006&   0.61&    0.71&   17.774&  18.133&   0.367&  ACEP-009\\
V03& 	04:59:00.4&  $-$67:44:40.3&   RRab&      	0.62198&   0.83&    1.17&   19.448&  19.956&   0.528&  RRLYR-02661\\
V04& 	04:59:01.1&  $-$67:43:29.4&   RRab&      	0.55585&   1.13&    1.36&   19.202&  19.559&   0.391&  RRLYR-02662\\
V05& 	04:59:02.8&  $-$67:46:06.0&   RRab& 	        0.69706&   0.59&    0.80&   19.301&  19.726&   0.439&  RRLYR-02674\\
V06& 	04:59:07.9&  $-$67:45:29.4&   RRc& 		0.30418&   0.60&    0.71&   19.449&  19.760&   0.312&  RRLYR-02712\\
V07& 	04:59:08.9&  $-$67:44:08.0&   RRc&       	0.29536&   0.56&    0.75&   19.292&  19.603&   0.325&  RRLYR-02727\\
V08& 	04:59:12.0&  $-$67:43:52.6&   RRab&      	0.57183&   0.91&    1.17&   19.227&  19.661&   0.456&  RRLYR-02748\\
V09& 	04:59:12.5&  $-$67:44:15.0&   RRc&       	0.36388&   0.47&    0.56&   19.445&  19.571&   0.132&  RRLYR-02757\\
V10& 	04:59:14.1&  $-$67:44:23.7&   RRab&      	0.69651&   0.69&    0.79&   19.474&  19.950&   0.480&  RRLYR-02770\\
V11& 	04:59:14.8&  $-$67:44:21.3&   RRc&       	0.31536&   0.54&    0.63&   19.339&  19.608&   0.275&  RRLYR-02777\\
V13& 	04:59:08.9&  $-$67:44:39.6&   RRab&    	0.55944&   0.36&    0.68&   17.761&  18.500&   0.746& - \\
V14& 	04:59:03.1&  $-$67:44:39.1&   RRc&       	0.31786&   0.62&    0.77&   19.473&  19.806&   0.345&  RRLYR-02676\\
V15& 	04:59:09.4&  $-$67:44:39.0&   RRab&   	0.74962&   0.35	 &    - &    18.26 &      - &       - &	  RRLYR-02732\\
V17& 	04:59:06.3&  $-$67:44:38.3&   RRab&   	0.52470&   0.33&    0.68&   17.855&  18.693&   0.852&  RRLYR-02697\\
V18& 	04:59:11.4&  $-$67:44:36.3&   RRab&      	0.64515&   0.97&    1.20&   19.083&  19.491&   0.439&  RRLYR-02747\\
V19& 	04:59:04.2&  $-$67:44:35.6&   RRab&   	0.74479&   0.43&    0.57&   18.807&  19.322&   0.522&  RRLYR-02680\\
V20& 	04:59:09.4&  $-$67:44:34.4&   ?& 	       	0.29193&   - 	 &    0.43&   - &      19.215&   - & -	\\
V22& 	04:59:10.1&  $-$67:44:27.2&   RRc&       	0.32920&   0.63&    0.72&   19.330&  19.564&   0.243&  RRLYR-02738\\
V23& 	04:59:04.8&  $-$67:44:24.9&   RRab&   	0.52851&   0.50&    0.93&   18.383&  19.030&   0.672&  RRLYR-02683\\
V24& 	04:59:11.2&  $-$67:44:17.4&   RRc&   	0.41110&   0.30&    0.39&   18.682&  18.955&   0.275&  RRLYR-02743/02746\\
V25-Field& 	04:59:41.3&  $-$67:44:16.8&   Ceph&     	3.27005&   0.84&    1.16&   15.694&  16.258&   0.601&  CEP-0591\\
V26& 	04:59:09.2&  $-$67:44:14.3&   RRab&      	0.73891&   0.39&    0.48&   19.179&  19.630&   0.455&  RRLYR-02729\\
V27& 	04:59:14.5&  $-$67:43:52.2&   RRab&      	0.66428&   0.58&    0.60&   18.990&  19.234&   0.245&  RRLYR-02776\\
V30& 	04:59:07.2&  $-$67:43:30.9&   Ceph&     	11.6769&  0.93&    1.55&   14.236&  14.799&   0.607&  CEP-0561\\
V31& 	04:59:06.6&  $-$67:43:26.2&   Eclipse& 	0.61/1.2&   - 	 &    - &      18.340&  18.376&   0.036&- \\
V32-Field& 	04:59:34.7&  $-$67:42:46.7&   RRab&      	0.532071&   1.11&    1.46&   19.089&  19.373&   0.325&  RRLYR-02855\\
V33 BL& 	04:59:28.9&  $-$67:42:40.1&   RRab&      	0.53278&   0.67&    0.71&   19.587&  20.028&   0.430&  RRLYR-02834\\
V34& 	04:58:59.3&  $-$67:42:38.3&   Eclipse& 	1.6/3.2 &   - 	 &    - &      19.514&  19.754&   0.240& -\\
V35-Field& 	04:59:36.0&  $-$67:42:31.9&   RRc&       	0.33058&   0.48&    0.57&   19.190&  19.556&   0.371&  RRLYR-02864\\
V36-Field& 	04:59:39.4&  $-$67:42:15.1&   Delta Scuti& 	0.07723&   0.56&    0.62&   20.319&  20.485&   0.172&  DSCT-400\\
V37& 	04:59:09.0&  $-$67:46:14.9&   RRc& 		0.35488&   0.44&    0.57&   19.353&  19.584&   0.239&  RRLYR-02728\\
V38& 	04:59:21.2&  $-$67:46:06.7&   RRab& 		0.75890&   0.40&    0.47&   19.014&  19.438&   0.427&  RRLYR-02802\\
V39& 	04:59:03.0&  $-$67:45:36.2&   Delta Scuti& 	0.08040&   0.28&    0.41&   20.699&  21.044&   0.349& -\\
V40& 	04:58:59.6&  $-$67:45:24.6&   RRc& 		0.33496&   0.55&    0.66&   19.631&  20.026&   0.403&  RRLYR-02657\\
V41& 	04:59:06.1&  $-$67:45:24.4&   T2Ceph& 		1.108051&   0.86&    1.07&   18.488&  18.928&   0.461&  T2CEP-020\\
V42& 	04:59:01.4&  $-$67:45:21.5&   RRc& 		0.37064&   0.45&    0.58&   19.307&  19.624&   0.325&  RRLYR-02667\\
V43& 	04:58:49.9&  $-$67:45:10.7&   Delta Scuti &	0.06096&   0.70&    0.84&   21.365&  21.636&   0.283& \\
V44& 	04:59:15.6&  $-$67:45:02.9&   RRc& 		0.36459&   0.51&    0.58&   19.294&  19.640&   0.348&  RRLYR-02781\\
V45& 	04:59:08.8&  $-$67:45:01.9&   RRc& 		0.29913&   0.68&    0.71&   19.389&  19.560&   0.180&  RRLYR-02724\\
V46& 	04:59:09.2&  $-$67:45:01.6&   Ceph& 		2.16488&   0.30&    0.41&   15.817&  16.491&   0.679&  CEP-0563\\
V49& 	04:59:05.7&  $-$67:44:55.9&   ? &		0.55403&   0.20&    0.24&   17.294&  18.223&   0.929& -\\
V50& 	04:59:10.0&  $-$67:44:55.6&   RRab&		0.49336&   0.74&    0.88&   18.601&  18.866&   0.263&  RRLYR-02737\\
V51& 	04:59:03.0&  $-$67:44:54.1&   RRab &		0.56363&   0.40&    0.58&   18.335&  18.898&   0.574&  RRLYR-02675\\
V52& 	04:59:09.3&  $-$67:44:53.4&   RRc &		0.32653&   0.47&    0.57&   18.989&  19.317&   0.333&  RRLYR-02730\\
V54& 	04:59:05.0&  $-$67:44:51.8&   RRc &		0.30898&   0.41&    0.48&   19.332&  19.669&   0.339&  RRLYR-02686\\
V55 BL& 	04:59:08.9&  $-$67:44:51.7&   RRab& 		0.51904&   0.97&    1.14&   18.720&  18.860&   0.154&  RRLYR-02725\\
V56& 	04:59:04.0&  $-$67:44:50.7&   RRab &		0.52757&   0.44&    0.67&   19.736&  20.036&   0.318&  RRLYR-02678\\
V57& 	04:59:09.5&  $-$67:44:50.5&   RRab &		0.52083&   - 	 &    - &      - &      - &       - &	  RRLYR-02733\\
V58& 	04:59:04.9&  $-$67:44:49.1&   RRab& 		0.55049&   0.63&    0.712&   18.821&  19.185&   0.371&  RRLYR-02685\\
V59& 	04:59:07.3&  $-$67:44:48.7&   RRab& 		0.56507&   - 	 &    - &      - &      - &       - &	  RRLYR-02706\\
V60& 	04:59:11.6&  $-$67:44:46.8&   RRc& 		0.29205&   0.32&    0.42&   19.217&  19.502&   0.289& -\\
V61& 	04:59:12.0&  $-$67:44:43.9&   RRc& 		0.36061&   0.44&    0.60&   19.302&  19.896&   0.597&  RRLYR-02749\\
V62& 	04:59:05.9&  $-$67:44:43.3&   RRc& 		0.36411&   0.28&    0.44&   18.969&  19.512&   0.550&  RRLYR-02692\\
V63& 	04:59:07.9&  $-$67:44:48.6&   RRab& 		0.62312&   - 	 &    - &      - &      - &       - &     RRLYR-02714\\
V64& 	04:59:07.8&  $-$67:44:52.6&   RRab& 		0.70228&   - 	 &    - &      - &      - &       - &     RRLYR-02711\\
V65& 	04:59:06.6&  $-$67:44:42.0&   RRab& 		0.64660&   - 	 &    - &      - &      - &       - &     RRLYR-02702\\
V66& 	04:59:07.5&  $-$67:44:35.7&   RRab& 		0.50999&   - 	 &    - &      - &      - &       - &     RRLYR-02671\\
\enddata
\label{1786vartable}
\end{deluxetable*}

\begin{figure*}
\epsscale{0.7}
\plotone{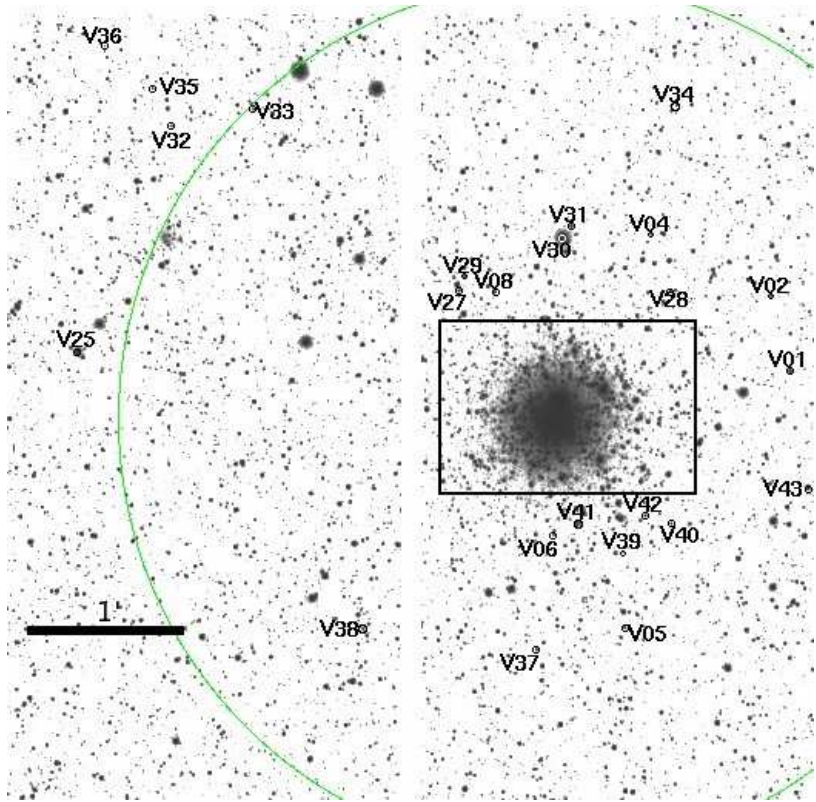}
\caption{Finding chart for the variable stars in the outer portion of NGC 1786.  North is up and East is to the left.  The white gap is due to the finding chart being made from a SOAR image and represents the $7.8$ arcsec mounting gap between the two CCDs of the SOI camera.  The large green circle shows the tidal radius of the cluster while the black box shows the region that is shown in Figure \ref{1786findin}.}
\label{1786findout}
\end{figure*}

\begin{figure*}
\epsscale{0.7}
\plotone{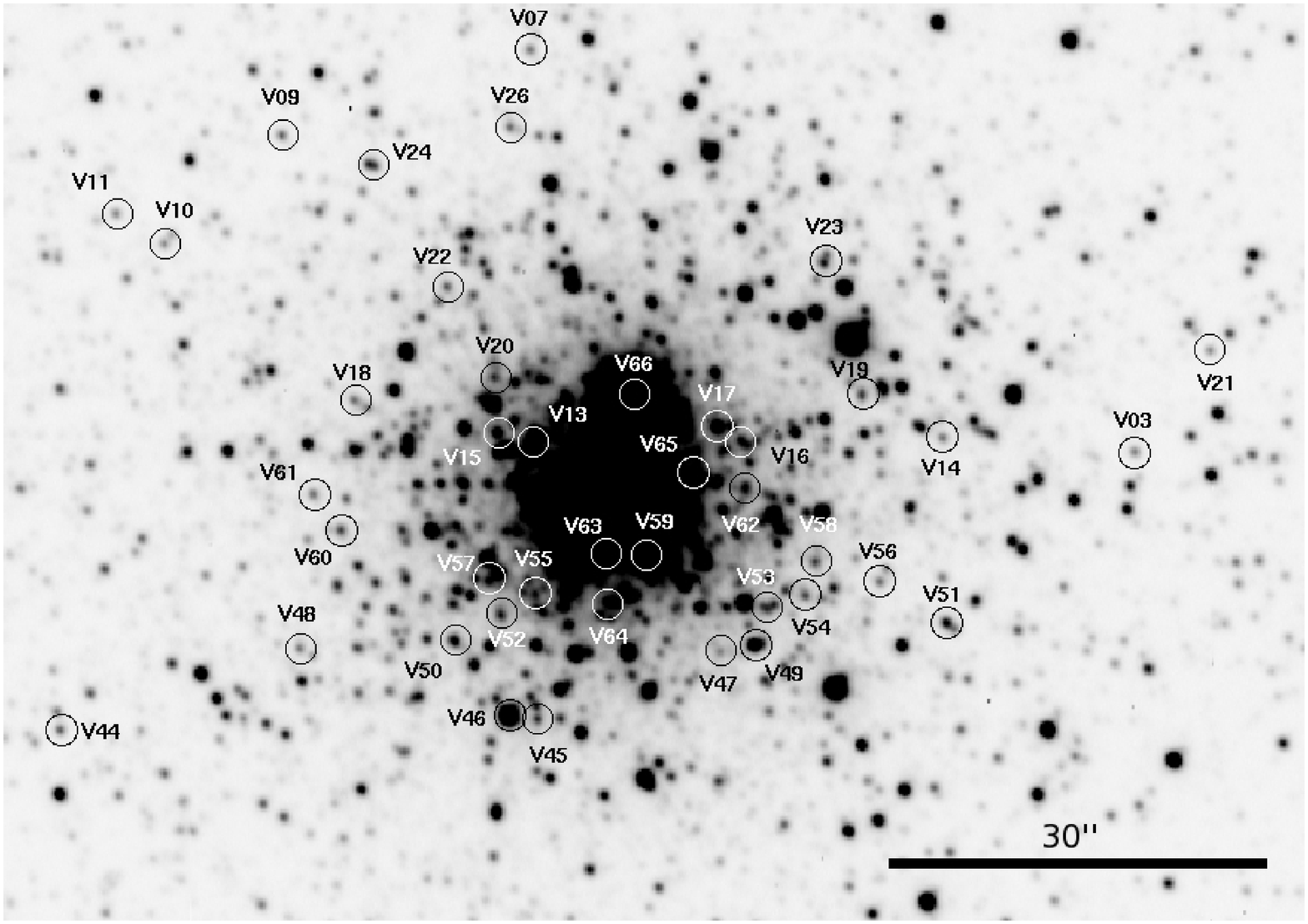}
\caption{Finding chart for the variable stars in the inner portion of NGC 1786.  North is up and East is to the left.}
\label{1786findin}
\end{figure*}

A total of $65$ variable stars were found within the SOAR $5.2 \times 5.2$ arcminute$^2$ field of view in the direction of NGC 1786.  This includes the cluster as well as the surrounding area.  The variables include $53$ RR Lyraes ($27$ RRab, $18$ RRc, and $8$ RRd), $3$ classical Cepheids, $1$ Type II Cepheid, $1$ Anomalous Cepheid, $2$ eclipsing binaries, $3$ Delta Scuti/SX Phoenicis variables, and $2$ variables of undetermined type.  The extreme crowding in the central region of NGC 1786 posed some difficulties for the identification and classification of the variable stars due to many of the stars being blended with near neighbors.  Table \ref{1786vartable} lists the identified variables, with the exception of the RRd stars which are listed in Table \ref{1786rrdtable}, their classification, period $V$ and $B$ amplitudes, intensity-weighted mean magnitudes, and magnitude-weighted mean $B-V$ color.  Variable stars V$01$ through V$11$ were originally found by \citet{graham85} and have periods confirmed by \citet{walker88}.  The names for these stars are the same as the ones used by Graham and Walker \& Mack.  The positions of the variable stars are shown in Figures \ref{1786findout} and \ref{1786findin}.

\begin{deluxetable*}{lcccccccccccccc}
\tablewidth{0pc}
\tabletypesize{\scriptsize}
\tablecaption{Photometric Parameters for the RRd Variables in NGC 1786}
\tablehead{\colhead{ID}& \colhead{RA} & \colhead{DEC} & \colhead{Type}& \colhead{$P_{0}$ (days)}& \colhead{$P_{1}$ (days)}&\colhead{$P_{1}/P_{0}$}& \colhead{$A_{V,0}$}& \colhead{$A_{V,1}$}& \colhead{$A_{B,0}$}& \colhead{$A_{B,1}$}& \colhead{$\langle V\rangle $}& \colhead{$\langle B\rangle $}& \colhead{$\langle B-V\rangle $} & \colhead{OGLE-III ID}}
\startdata
V02& 	04:58:52.6&  $-$67:43:53.8&   RRd&   0.4876&  	0.3630 	&  0.7445&  0.24&  0.38&  0.20&  0.50&   19.144&  19.442&   0.306&   RRLYR-02626\\
V16&	04:59:06.0&  $-$67:44:39.6&   RRd&   	0.4838& 0.3601& 0.7443 & - &  - 	 &    - &      - &      - &    -&   - &	  RRLYR-02693\\
V21& 	04:58:59.3&  $-$67:44:32.1&   RRd&   0.4937&  	0.3676&  0.7447&  0.16&  0.46&  0.22&	 0.58&   19.443&  19.806&   0.367&   RRLYR-02653\\
V28& 	04:58:59.7&  $-$67:43:52.1&   RRd&   0.4876&  	0.3631 &   0.7447&  0.28&  0.44&  0.32&  0.62&   19.200&  19.493&   0.308&   RRLYR-02658\\
V29& 	04:59:14.1&  $-$67:43:46.2&   RRd&       0.4955&	0.3694&  0.7455&  -& 0.26& -&   0.38&   19.083&  19.523&   0.444&  RRLYR-02771\\
V47& 	04:59:06.2&  $-$67:44:56.3&   RRd& 	0.4969&	0.3699&  0.7444& - & 0.75&   - & 1.00&   19.977&  20.444&   0.491&  RRLYR-02698\\

V48& 	04:59:12.2&  $-$67:44:56.3&   RRd&   0.4801&	0.3574 &  0.7445& 0.28&  0.36&  0.46&  0.36&   19.324&  19.729&   0.409&   RRLYR-02750\\
V53& 	04:59:05.6&  $-$67:44:52.8&   RRd &	0.4918&	0.3661&  0.7444&  - & - & -	 &    - &      - &      - &       - &	  RRLYR-02688\\

\enddata
\label{1786rrdtable}
\end{deluxetable*}

\subsection{Field Contamination}

\begin{figure}
\epsscale{1.2}
\plotone{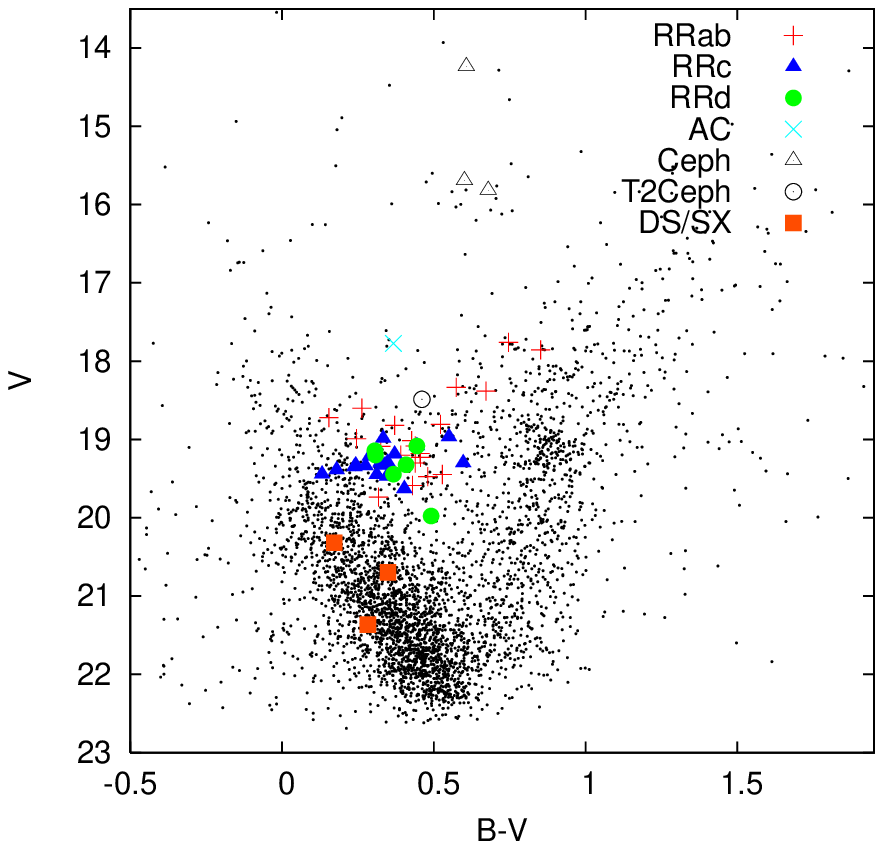}
\caption{$V$,$B-V$ CMD for stars within the tidal radius of NGC 1786 with the positions of the RR Lyrae variables also indicated.  Plus symbols indicate RRab stars, filled triangles indicate RRc's, and circles indicate RRd's.  The contamination by LMC field stars is readily apparent.  The scatter in the positions of the RR Lyrae stars is a consequence of blending that arises from the crowded nature of the central region of the cluster.}
\label{1786cmd}
\end{figure}

\begin{figure}
\epsscale{1.2}
\plotone{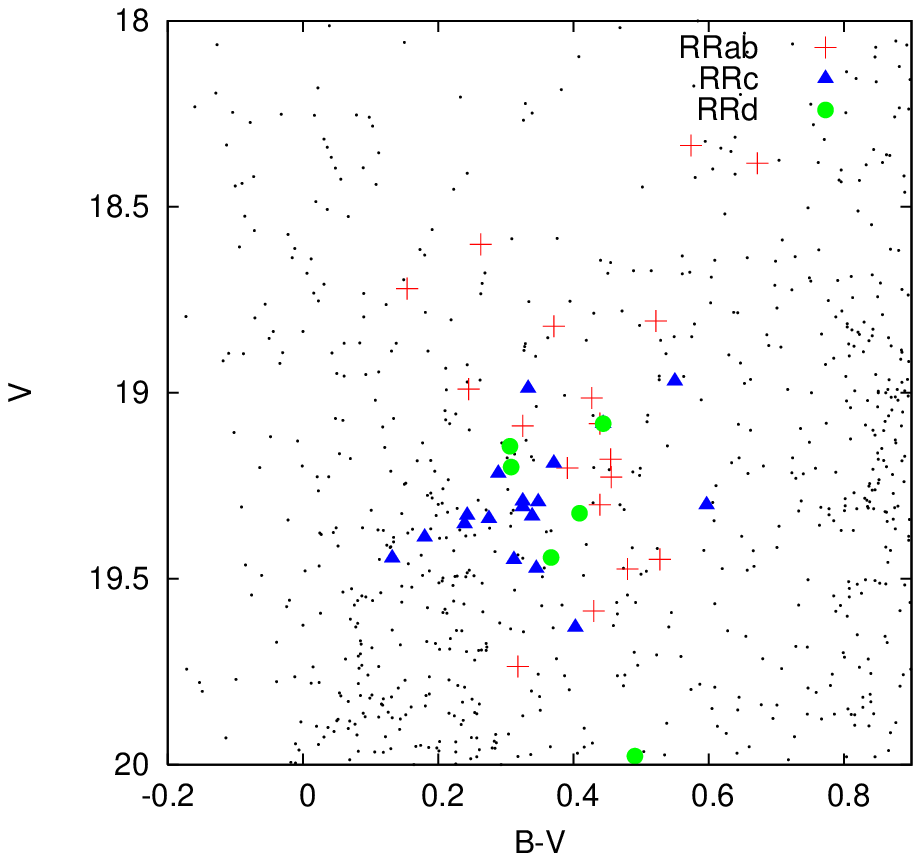}
\caption{$V$,$B-V$ CMD for NGC 1786 that is zoomed in on the horizontal branch.  The symbols used are the same as in Figure \ref{1786cmd}}
\label{1786cmdzoom}
\end{figure}

NGC 1786's location toward the center of the LMC means that a significant amount of contamination from LMC field stars is expected in the images.  This makes it difficult to determine with absolute certainty whether an individual variable star is a member of the cluster.  NGC 1786 has a tidal radius of $2.9$ arcminutes \citep{mclaughlin05}, and $61$ of the $65$ variable stars are located within that tidal radius; the tidal radius does extend beyond the western edge of our field, and slightly beyond the northern and southern edges.  The $4$ stars ($1$ RRab, $1$ RRc, $1$ classical Cepheid, and $1$ Delta Scuti/SX Phoenicis) outside the tidal radius are not cluster members and they are noted in Table \ref{1786vartable} with the word ``Field'' after their name.  The field of view in the images obtained with SOAR is $5.2 \times 5.2$ arcminutes$^2$ and a visual inspection shows that roughly $75\%$ of the image is contained within the tidal radius of NGC 1786.  The LMC field should feature a roughly even distribution of variable stars, thus the majority of the $61$ variables found within the tidal radius should be members of the cluster.  Classical Cepheids are younger stars, thus the two classical Cepheids that are within the tidal radius of NGC 1786 should be members of the LMC field.  The amount of LMC field star contamination is still significant within the tidal radius of the cluster, as can be seen in Figures \ref{1786cmd} and \ref{1786cmdzoom}, where the presence of the field stars makes it difficult to distinguish a horizontal branch in the CMD for NGC 1786.  The scatter in the position of the RR Lyrae stars in the CMD is likely due to the crowded nature of the central region of the cluster, which results in many of the RR Lyrae being partially blended with nearby stars, affecting their brightness and color.

The OGLE-III database \citep{soszy09} can also be used to get a better understanding of the amount of LMC field star contamination within the tidal radius of NGC 1786.  We looked at an annulus centered on NGC 1786 which had the same area on the sky as contained within the tidal radius of the cluster, but which had a radius greater than the tidal radius.  Within that annulus OGLE-III found $7$ RRab, $2$ RRc, $1$ classical Cephied, and $1$ Delta Scuti star.  Within the $2.9$ arcminute tidal radius OGLE-III found $31$ RRab, $19$ RRc, $9$ RRd, $1$ Type II Cepheid, $2$ classical Cepheids, $1$ Anomalous Cepheid.  As previously mentioned, the classical Cepheids cannot be cluster members due to their ages and the numbers of Type II and Anomalous Cepheids are too small to be able to say anything about the likelihood of their membership in the cluster in a statistical way.  Since there is no reason to suspect a significant variation in the distribution of LMC field RR Lyrae stars over the size of the fields in question, it is reasonable to suspect that approximately $15\%$ of the RR Lyrae stars found within the tidal radius of NGC 1786 to actually be members of the LMC field.  It is interesting to note that the a large majority of the RR Lyrae stars found in the comparison field are RRab stars and none of them are RRd stars; this is a marked contrast to what is found within the tidal radius where $32\%$ of the RR Lyrae are RRc and $15\%$ are RRd.  Based on this we should expect that the majority of LMC field RR Lyrae within the tidal radius are RRab stars and that the ratio of first overtone dominant RR Lyrae to total number of RR Lyrae among cluster members is higher than suggested by the numbers found within the tidal radius; this will be discussed in more detail later in this paper when we discuss the Oosterhoff classification of NGC 1786.

%\thesisfigure{image}{rotation}{scale}{caption text}{short caption text}{label}

\begin{deluxetable*}{lcccccc}
\tablewidth{0pc}
\tabletypesize{\scriptsize}
\tablecaption{Photometry of the Variable Stars}
\tablehead{\colhead{ID} & \colhead{Filter} & \colhead{HJD} & \colhead{Phase} & \colhead{Mag} & \colhead{Mag Error} & \colhead{Telescope} }
\startdata
V01 & $B$ & 2453981.8230 & 0.11033 & 17.878 & 0.013 & SMARTS\\
V01 & $B$ & 2453988.7630 & 0.78468 & 18.230 & 0.070 & SMARTS\\
V01 & $B$ & 2453991.7793 & 0.55477 & 18.703 & 0.063 & SMARTS\\
V01 & $B$ & 2453993.8316 & 0.11995 & 17.971 & 0.015 & SMARTS\\
V01 & $B$ & 2453995.8099 & 0.59264 & 18.564 & 0.021 & SMARTS\\
\enddata
\tablecomments{Maximum light occurs at a phase of $0$.  This table is published in its entirety in the electronic edition.}
\label{phottable}

\end{deluxetable*}

\subsection{RR Lyrae Stars}

\begin{figure*}
\begin{center}
\includegraphics[width=0.45\textwidth]{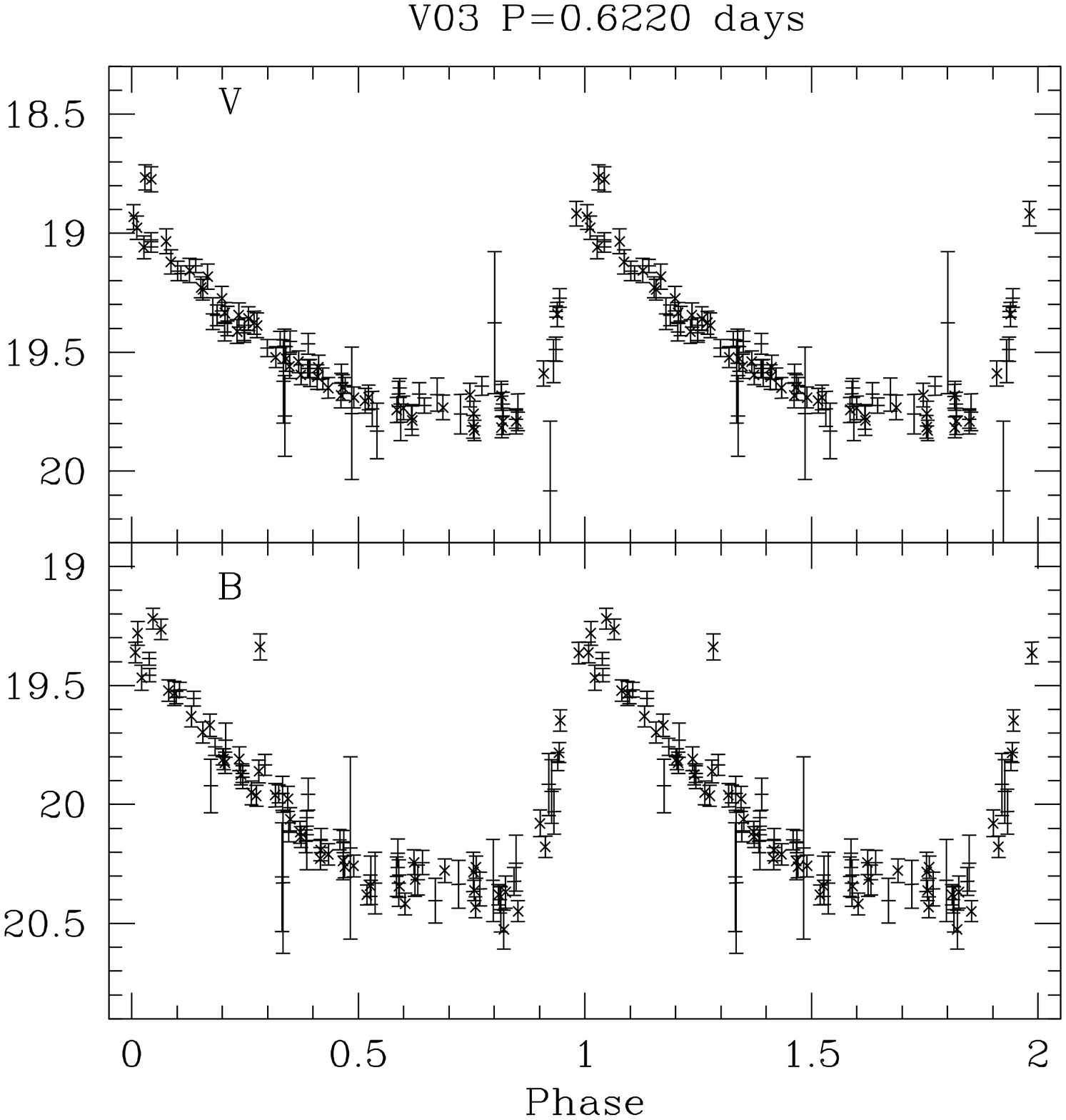}
\includegraphics[width=0.45\textwidth]{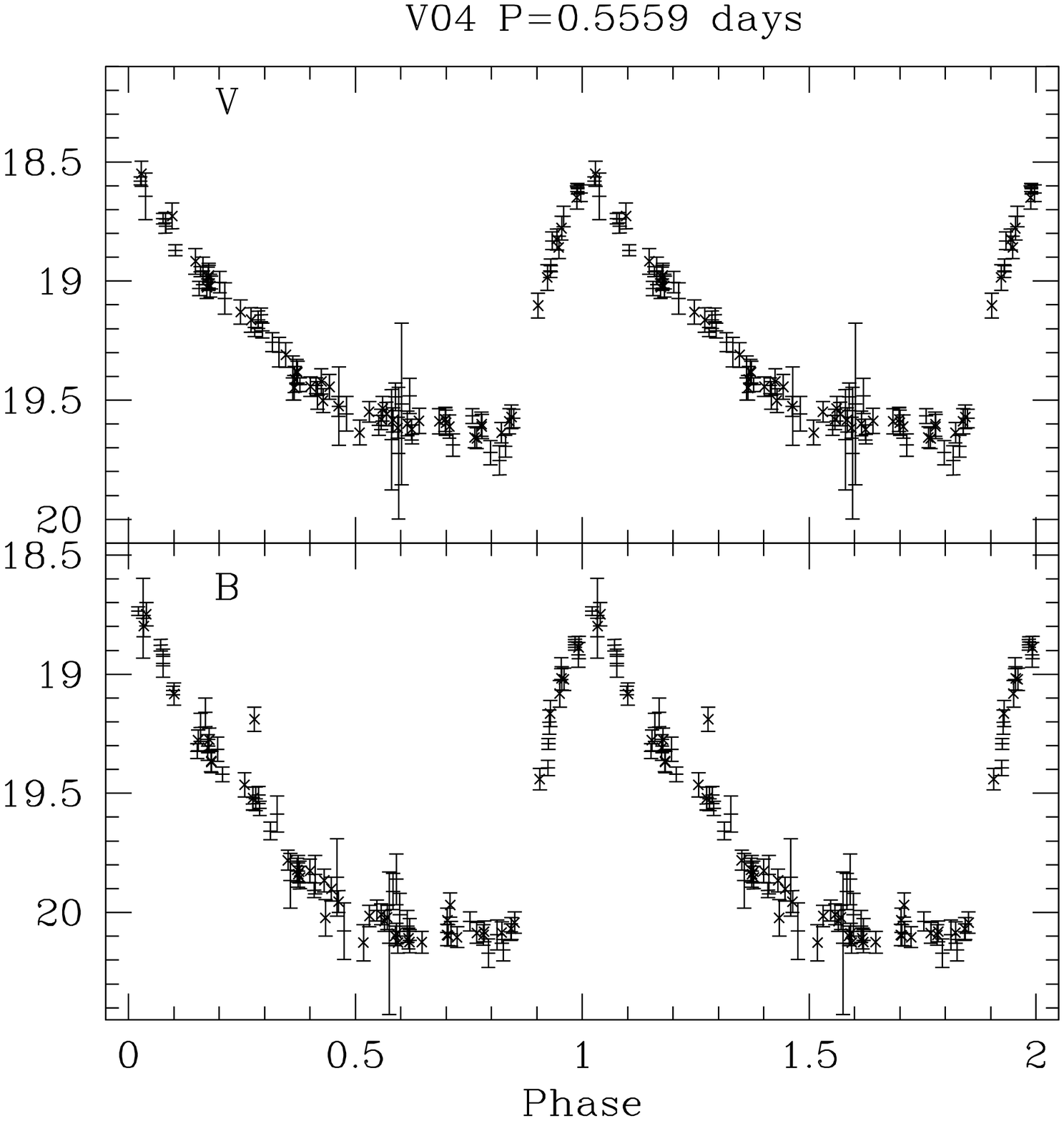}
\caption{Sample light curves for RRab stars in NGC 1786. SOAR data is indicated with crosses while SMARTS data is indicated by points with horizontal lines.  (The full set of light curves can be found in the electronic verson of this paper.)}
\label{1786ab}
\end{center}
\end{figure*}

\begin{figure*}
\begin{center}
\includegraphics[width=0.45\textwidth]{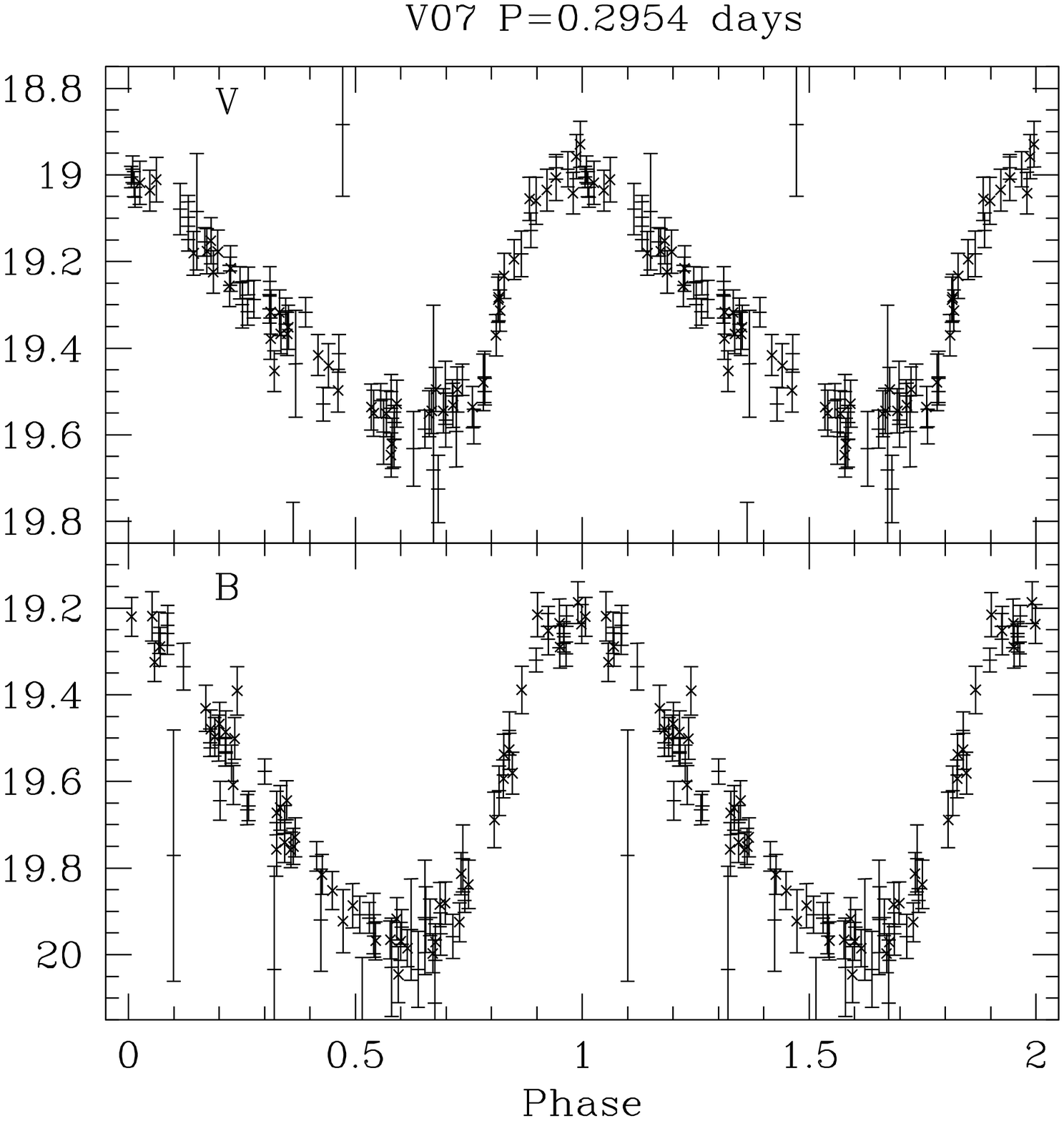}
\includegraphics[width=0.45\textwidth]{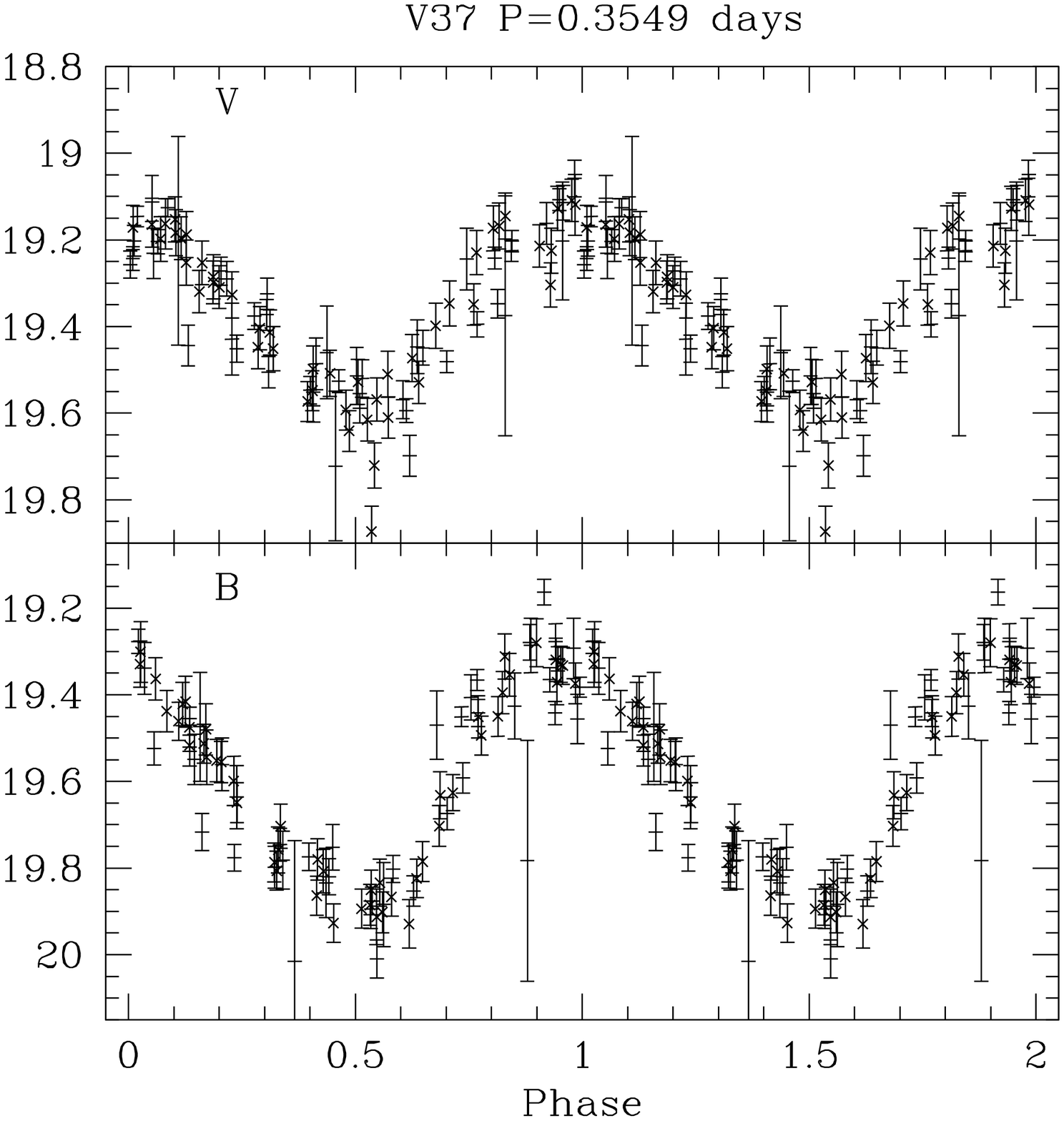}
\caption{Sample light curves for RRc stars in NGC 1786. SOAR data is indicated with crosses while SMARTS data is indicated by points with horizontal lines. (The full set of light curves can be found in the electronic verson of this paper.)}
\label{1786c}
\end{center}
\end{figure*}

\begin{figure*}
\begin{center}
\includegraphics[width=0.45\textwidth]{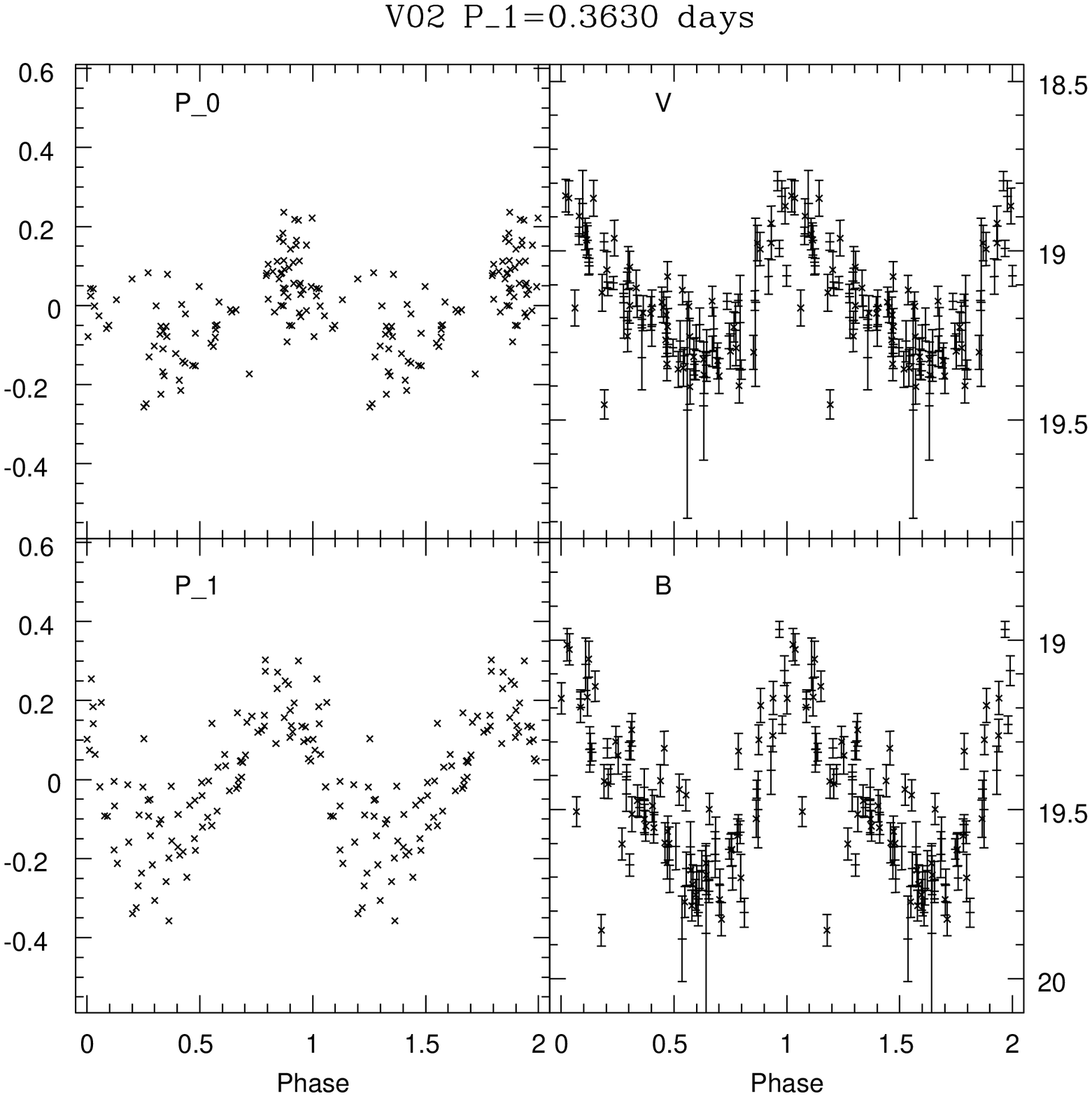}
\includegraphics[width=0.45\textwidth]{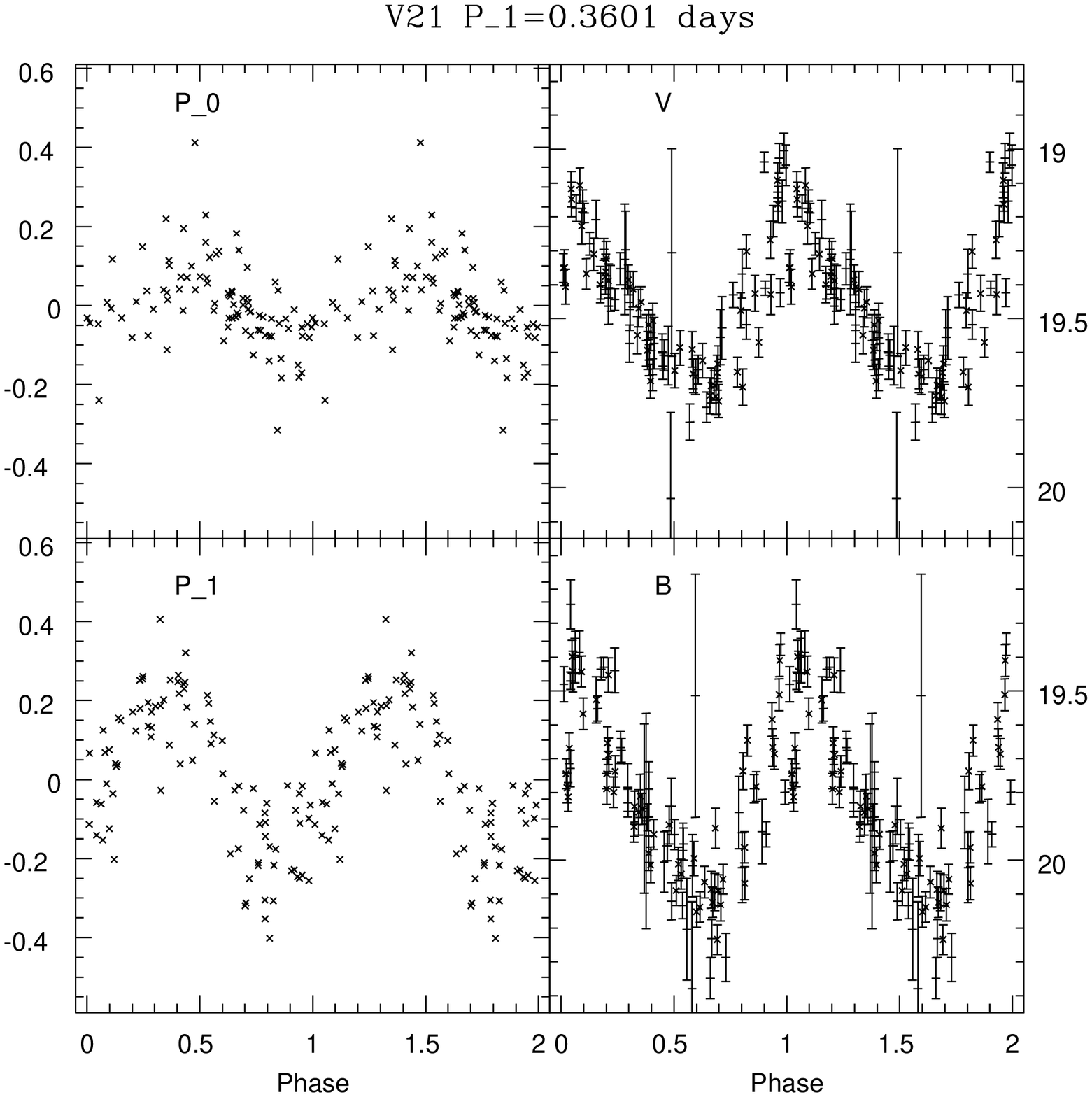}
\caption{Sample light curves for RRd stars in NGC 1786. The left panels show the residuals left from subtracting the fundamental or first overtone periods.  The right panels show the $V$ and $B$-band light curves plotted with the first overtone period.  SOAR data is indicated with crosses while SMARTS data is indicated by points with horizontal lines. (The full set of light curves can be found in the electronic verson of this paper.)}
\label{1786d}
\end{center}
\end{figure*}

\begin{figure*}
\begin{center}
\includegraphics[width=0.45\textwidth]{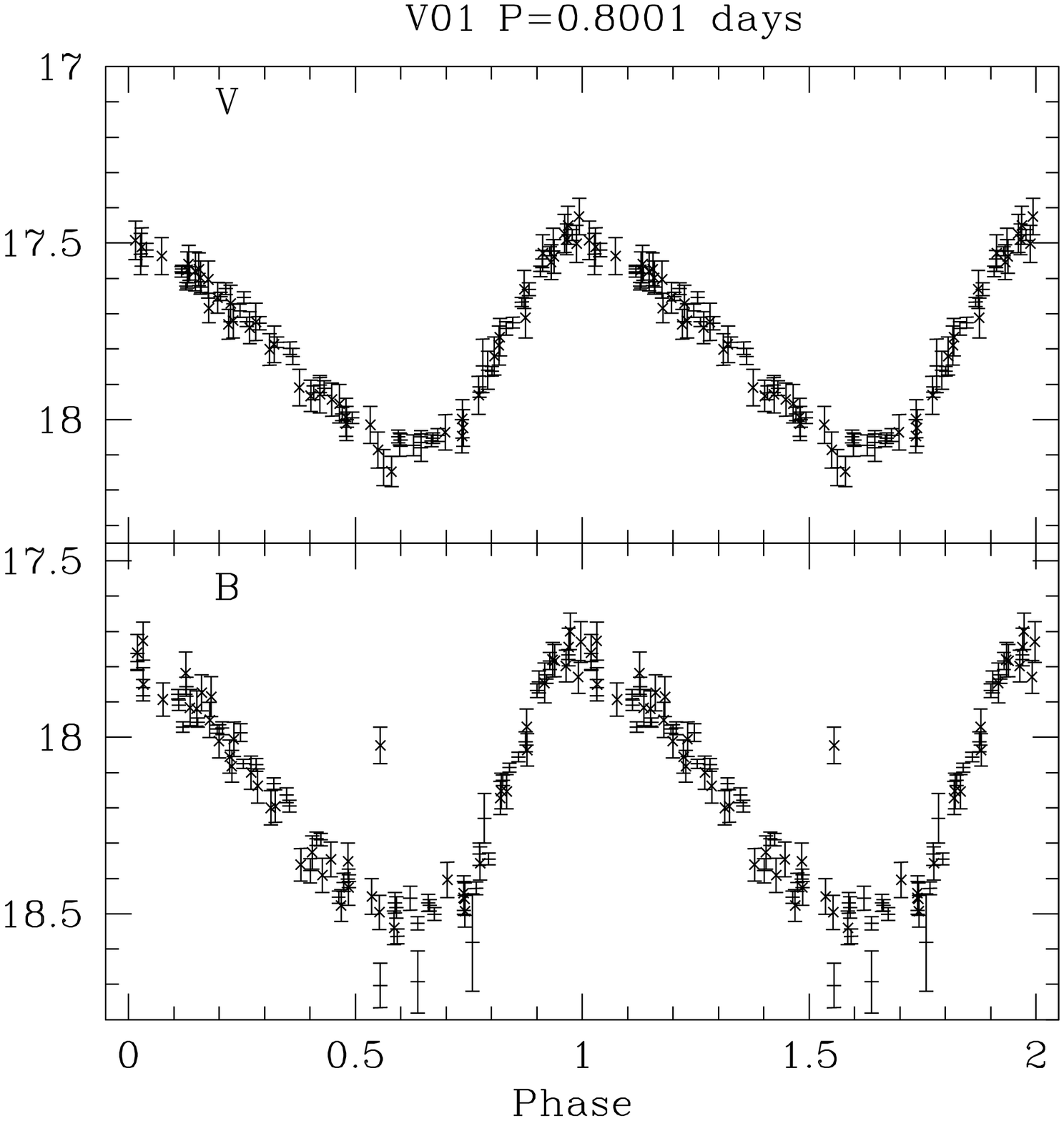}
\includegraphics[width=0.45\textwidth]{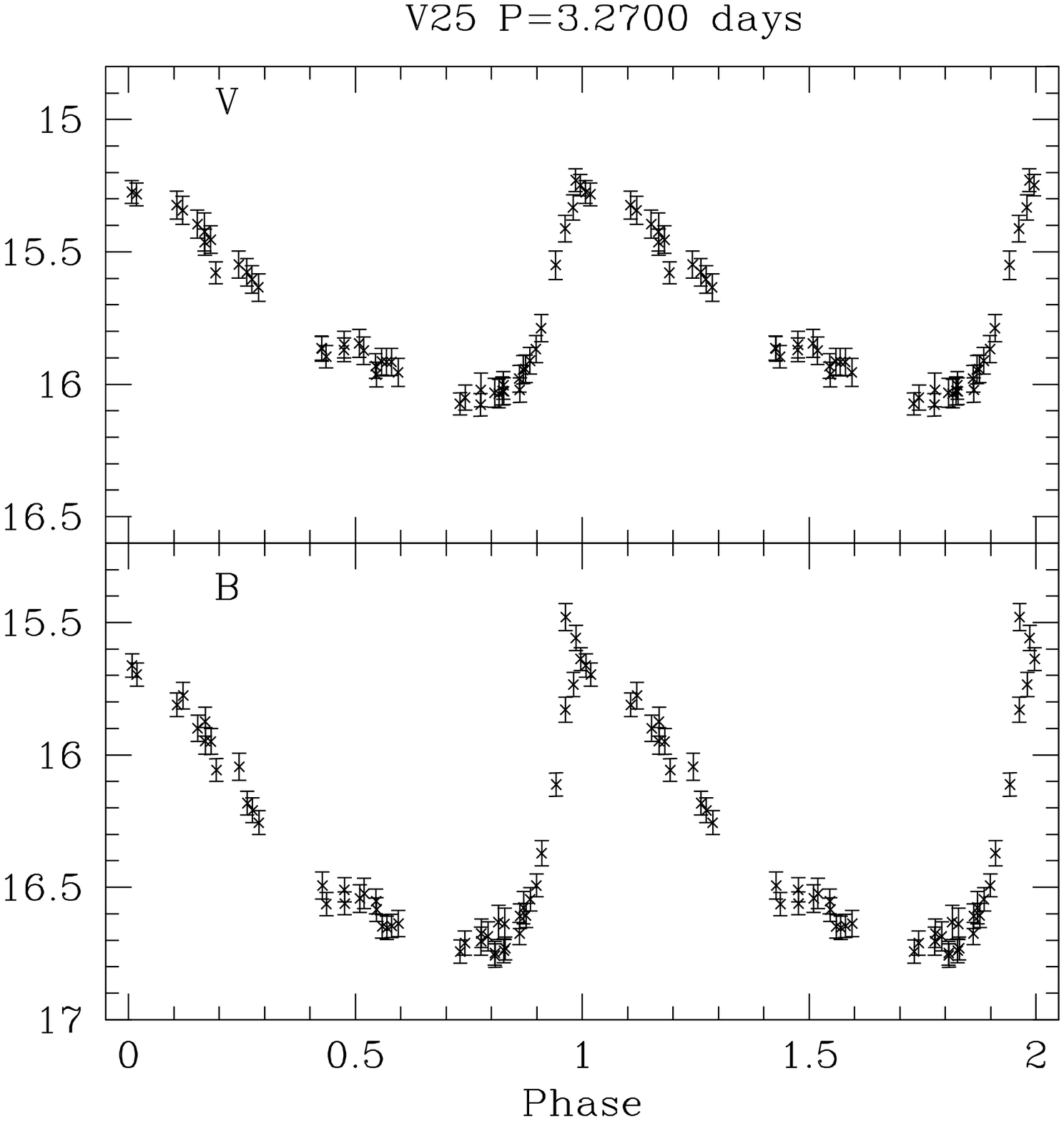}
\includegraphics[width=0.45\textwidth]{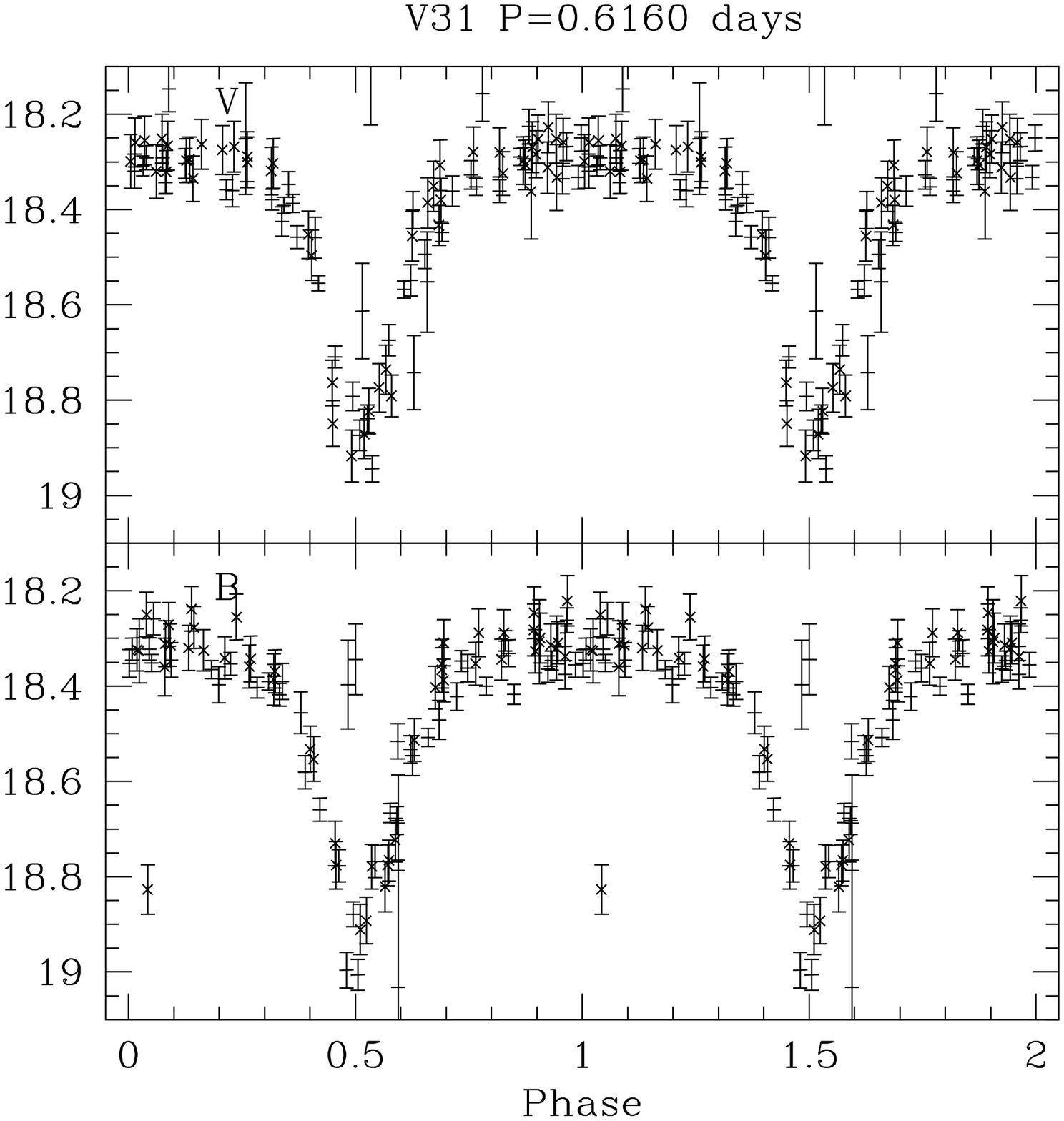}
\includegraphics[width=0.45\textwidth]{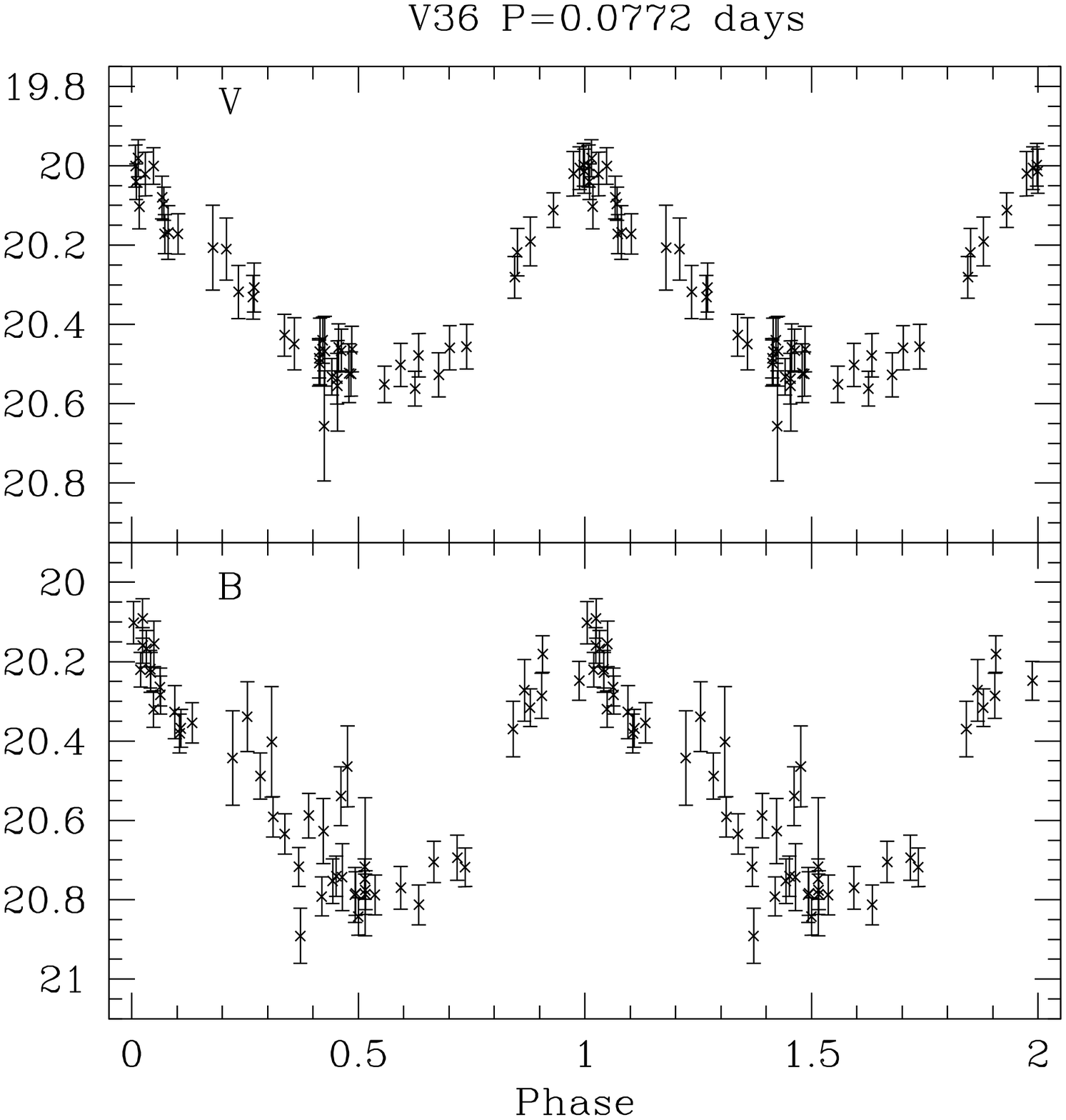}
\caption{Sample light curves for the non-RR Lyrae variable stars in NGC 1786. V$01$ is an Anomalous Cepheid while V$25$ is a classical Cepheid.  The second row shows the light curves of one of the eclipsing binaries and one of the Delta Scuti/SX Phoenicis stars.  SOAR data is indicated with crosses while SMARTS data is indicated by points with horizontal lines. (The full set of light curves can be found in the electronic verson of this paper.)}
\label{1786othervar}
\end{center}
\end{figure*}

A total of $26$ RRab, $17$ RRc, and $8$ RRd stars were found within the tidal radius of NGC 1786; $1$ RRab and $1$ RRc are new discoveries that have not been found in any of the previous studies of the cluster.  All of the $9$ RR Lyrae stars originally identified by \citet{graham85} were found.  V$2$ had been classified as an RRc star by \citet{graham85} and \citet{walker88}.  However, the larger number of epochs in our data set allowed for the detection of a second pulsation mode in this star, and it has accordingly been reclassified as an RRd star.  For these nine stars originally found by Graham, we obtained intensity-weighted mean magnitudes of $\langle V\rangle=19.321\pm0.039$ and $\langle B\rangle=19.677\pm0.059$.  These are fainter than Walker \& Mack's values ($\langle V\rangle=19.203\pm0.073$ and $\langle B\rangle=19.526\pm0.066$).  This disagreement in brightness is potentially due to the improved resolution of our observations, decreasing the effect of blending between the RR Lyrae stars and other nearby stars.  V$08$ shows strong indications of being blended in Walker \& Mack's data, appearing brighter than the other RR Lyrae, while in our data it has a similar brightness to the other RR Lyrae.  We also found large amplitudes for V$08$ in our data than were obtained by Walker \& Macks ; this is expected for a blended star.  If we remove V$08$ from the calculation, the mean magnitudes for the RR Lyrae from Walker \& Mack become $\langle V\rangle=19.271\pm0.030$ and $\langle B\rangle=19.573\pm0.052$; these are much closer to the values we obtained.  For the entire set of RR Lyrae stars within the tidal radius found in our study, the mean magnitudes are $\langle V\rangle=19.241\pm0.047$ and $\langle V\rangle=19.594\pm0.052$; these values are brighter than our values for just the stars that were originally found by Graham. Since the majority of the new RR Lyrae were found toward the crowded cluster center, one would expect that they would more likely suffer from blending and thus appear brighter, this is discussed in Section 5.  Sample light curves for the RR Lyrae stars in NGC 1786 are shown in Figures \ref{1786ab}-\ref{1786othervar} with the SOAR data shown indicated by crosses and the SMARTS data by points with a horizontal line; the full set of light curves can be found in the electronic version of this paper.  Table \ref{phottable} contains the photometric data for the variable stars.

\citet{soszy09} found $57$ RR Lyrae stars within the tidal radius of NGC 1786 using the OGLE-III data.  We found $51$ of these stars in our data; the remaining six are located within the extremely crowded core of the cluster.  These six stars were only found in the OGLE $I$-band data, likely due to the seeing be better in $I$ than in $B$ or $V$.
%Of the OGLE-III stars that were found in our data set, $33$ were classified as RR Lyrae stars ($13$ RRab, $16$ RRc, and $4$ RRd) and $16$ as candidate RR Lyrae.  

%V$06$ (OGLE-III RRLYR-02712) was classified as an RRab star by \citet{soszy09}, however we found it to have an intensity-weighted mean magnitude of $\langle V\rangle=20.539$, about one magnitude fainter than the horizontal branch.  Soszy\'{n}ski et al. found a mean magnitude of $\langle V\rangle=19.410$ for this star, about one magnitude brighter than our value.  Since V$06$ is not in a very crowded region of the cluster, the reason for this disagreement in brightness is unknown.

\citet{soszy09} found $9$ RRd stars, all of which show up in our data but only $8$ of which we classified as RRd stars.  V$16$ (OGLE-III RRLYR-02693), V$29$ (OGLE-III RRLYR-02771), V$47$ (OGLE-III RRLYR-02698), and V$53$ (OGLE-III RRLYR-02688) were all classified as an RRd stars by \citet{soszy09}; however, while they do display some possible double-mode behavior, we were unable to determine  second periods for these stars.  While our data set is larger in the $V$-band than the OGLE-III data set, OGLE-III's $I$-band has far more observation than either $V$-band data set and is thus better equipped for the determination of secondary periods in double-mode variables.  Our data does not contradict OGLE-III's results for these stars so we have classified these four stars as RRd.  The listed periods for these stars in Table \ref{1786rrdtable} are taken from OGLE-III; the lack of listed amplitudes for the fundamental mode for these stars comes from our inability to find reliable second periods for them.  

OGLE-III RR LYR-02746 was classified as an RRd star bu \citet{soszy09} but is only slightly seperated on the sky from OGLE-III RRLYR-02743 ($\approx0.15$ arcseconds) and these can potentially be a single star.  Our photometry only found one variable star at this location, V$24$, which has appears to be an RRc star, though the light curve is noisy.  We were unable to find any secondary periods for this star and it is likely that the noisy light curve is a result of blending with nearby non-variable neighbors; this would also explain the star being brighter than the horizontal branch.
%Several of these stars, V$16$ (OGLE-III RRLYR-02693), V$47$ (OGLE-III RRLYR-02698), and V$53$ (OGLE-III RRLYR-02688), were classified as candidate RR Lyrae stars for reasons that are discussed below.

V$48$ (OGLE-III RRLYR-02750) is an RRd star however it is questionable if it is a first overtone dominant RRd star or a fundamental-mode dominant pulsator.  The $V$-band light curve indicates the first overtone amplitude is larger, which agrees with the $I$-band results from \citet{soszy09}.  However, the $B$-band data suggests that the fundamental-mode amplitude is larger.  Since there are more $V$-band data points, we have decided to consider the first overtone dominant in V$48$ when it comes to determining the number ration of first overtone dominant pulsators to fundamental-mode pulsators.  This agrees with the results from OGLE-III's $I$-band data which indicated the first overtone was dominant.

V$13$, V$15$ (OGLE-III RRLYR-02732), V$17$ (OGLE-III RRLYR-02697), V$23$ (OGLE-III RRLYR-02683), and V$15$ (OGLE-III RRLYR-02675) are all RR Lyrae stars that are at least one magnitude brighter than the mean for the other RR Lyrae stars and have amplitudes that are smaller than expected for their period.  These stars are all located toward the center of the cluster and are likely blended; the difference in amplitude is exactly what you would expect from the amount of blending needed to increase their magnitude to the observed level.  Because of this blending we do not include these $5$ stars when calculating the average brightnesses of the NGC 1786 RR Lyraes; they are included in all mean period determinations as blending has no impact on the observed period.

V$33$ (OGLE-III RRLYR-02834) features a light curve that suggests that it may exhibit the Blazhko effect.  V$55$ (OGLE-III RRLYR-02725) also features a light curve that suggests a possible Blazhko effect but this star is a bit brighter than the other RR Lyrae stars and thus it is possible the noise in the light curve is a result of blending.  These stars are indicated as being possible Blazhko stars by the letters 'BL' in Table \ref{1786vartable}.

ISIS was used to find $9$ of the RR Lyrae stars (V$15$, V$16$, V$53$, V$57$, V$59$, V$63$-V$66$); the location of these stars in the extremely crowded center of the cluster did not allow Daophot light curves to be obtained.  

%Of the remaining candidate RR Lyrae stars, V$17$, V$23$, and V$51$ are more than one magnitude brighter than the horizontal branch and have colors that are redder than the confirmed RR Lyrae stars, possibly due to blending.  V$47$ (OGLE-III RRLYR-02698) is approximately $0.5$ magnitudes fainter than the horizontal branch and shows potential RRd behavior, but a second period could not be determined; OGLE-III classified it as an RRd star.  V$56$ has a large gap in its light curve which potentially impacts the accuracy of the determination of its period and amplitude.

\subsection{Other Variables}

In addition to the RR Lyrae stars, our field of view contained $3$ classical Cepheids, $1$ Type II Cepheid, $1$ Anomalous Cepheid, $2$ eclipsing binaries, and $3$ stars that are either Delta Scuti or SX Phoenicis variables.  These classifications were determined based on the period,brightness, and light curve shape; these classifications agree with those from OGLE-III and the literature.

V$25$, V$30$, and V$46$ are all classical Cepheids.  As classical Cepheids are much younger stars than RR Lyrae's, they should be members of the LMC field population.  The fact that two of the three classical Cepheids fall within the tidal radius of NGC 1786 on the image, is an indication that the LMC field population makes a significant contribution to the number of variable stars that fall in this region.

\begin{deluxetable*}{lccccccccc}
\tablewidth{0pc}
\tabletypesize{\scriptsize}
\tablecaption{Fourier Coefficients for RRab Variables}
\tablehead{\colhead{ID} & \colhead{$A_{1}$} & \colhead{$A_{21}$} & \colhead{$A_{31}$} & \colhead{$A_{41}$} & \colhead{$\phi_{21}$} & \colhead{$\phi_{31}$} & \colhead{$\phi_{41}$}& \colhead{$D_{max}$}& \colhead{Order}}
\startdata
V03&	0.369&	0.446&	0.249&	0.178&	2.47&	5.03$\pm$0.18&	1.25&	46.35& 6\\
V04&	0.454&	0.396&	0.232&	0.090&	2.30&	4.93$\pm$0.10&	1.09&	26.97& 10\\
V05&	0.229&	0.397&	0.318&	0.122&	2.26&	5.54$\pm$0.17&	2.34&	12.15& 7\\
V08&	0.308&	0.492&	0.320&	0.107&	2.13&	4.75$\pm$0.27&	1.20&	5.65& 7\\
V10&	0.235&	0.564&	0.300&	0.173&	2.39&	4.97$\pm$0.30&	1.76&	6.73& 7\\
V26&	0.175&	0.377&	0.153&	0.076&	2.46&	5.88$\pm$0.36&	2.27&	39.77& 8\\
V38&	0.161&	0.336&	0.192&	0.102&	2.96&	6.05$\pm$0.32&	2.32&	8.42& 7\\
\enddata
\label{1786abcoeff}
\end{deluxetable*}

Three short period variables, V$36$, V$39$, and V$43$, were found and have been classified as either Delta Scuti stars, which would be part of the LMC field, or SX Phoenicis stars, which would be member of NGC 1786.  \citet{soszy09} classified V$36$ (OGLE-III DSCT-400) as a Delta Scuit and its location outside of the tidal radius of NGC 1786, making it an LMC field star, is consistent with it most likely being a Delta Scuti.

Two eclipsing binaries, V$31$ and V$34$, were found.  Potential periods are listed in Table 1 but we did not have enough observations to definitively determine periods.

\section{Physical Properties of the RR Lyrae Stars}

Physical properties for the RR Lyrae stars were determined using the Fourier decomposition method.  The RRab light curves were fit with a Fourier series of the form
\begin{equation}
m(t)=A_{0}+\sum_{j=1}^{n}A_{j}\sin (j\omega t+\phi_{j}),
\end{equation}
while the RRc light curves were fit with a cosine series.  The resulting Fourier coefficients were then used to calculate physical properties of the stars using the relations from \citet{jk96}, \citet{ju98}, Kov\'acs \& Walker (1999, 2001), \citet{sc93}, and \citet{mw07}.  Full details of this method are discussed in the first paper in this series \citep{kuehn11}.

We attempted to fit Fourier series to the light curves of all the RR Lyrae stars but only $7$ RRab and $7$ RRc stars had light curves of sufficient quality to allow for a successful fit.  One of the RRc stars, V$35$, is located outside of the tidal radius of NGC 1786 and is thus a member of the LMC field, this is indicated in the tables.  Tables \ref{1786abcoeff} and \ref{1786ccoeff} list the Fourier coefficients for the RRab and RRc stars, respectively, while Tables \ref{1786abphys} and \ref{1786cphys} list the physical properties of these stars.  The averages of the physical properties for the RRc stars are calculated using only the $6$ stars within the tidal radius of the cluster.  Table \ref{1786abcoeff} also lists the $D_m$ value from \citet{jk96}.  Lower $D_m$ values indicate RRab stars with more ``regular'' light curves while higher values indicating those with more ``anomalous'' light curves.

The mean metallicity for the RRab stars is ${\rm [Fe/H]_{J95}}=-1.48\pm0.14$ in the scale of \citet{ju95}.  This can be transformed into the \citet{zw84} scale using the relation ${\rm [Fe/H]_{J95}} =1.431{\rm [Fe/H]_{ZW84}}+0.880$ \citep{ju95}.  This gives a mean metallicity for the RRab stars of ${\rm [Fe/H]_{ZW84}}=-1.65\pm0.10$.  This is more metal-rich than the literature values, though it is consistent within the error bars.  If we consider only the two stars with the smallest $D_{max}$ (V$08$ and V$10$), indicating that they have the most regular light curves, the average metallicity becomes ${\rm [Fe/H]_{ZW84}}=-1.96\pm0.13$.  This value is more metal poor than the literature values.  The mean metallicity found for the RRc stars is ${\rm [Fe/H]_{ZW84}}=-1.92\pm0.06$, which is more metal-poor than any of the literature values but is consistent with the value for the two RRab stars with the lowest $D_{max}$ values.  The LMC field RRc star, V$35$, in Table \ref{1786cphys} immediately stands out by being much more metal-rich than any of the other RRc stars.  In the new UVES scale \citep{carretta09} these metallicities are ${\rm [Fe/H]_{UVES}}=-2.04\pm0.19$ for the two RRab stars with the lowest $D_{max}$ values and ${\rm [Fe/H]_{UVES}}=-1.99\pm0.10$.

\begin{deluxetable*}{lcccccccc}[t]
\tablewidth{0pc}
\tabletypesize{\scriptsize}
\tablecaption{Fourier Coefficients for RRc Variables}
\tablehead{\colhead{ID}&  \colhead{$A_{1}$}& \colhead{$A_{21}$}& \colhead{$A_{31}$}& \colhead{$A_{41}$}& \colhead{$\phi_{21}$}& \colhead{$\phi_{31}$}& \colhead{$\phi_{41}$}& \colhead{Order}}
\startdata
V07&	   	0.274&	0.284&	0.121&	0.079&	 4.44&	 2.23$\pm$0.36&	1.22&	 8\\
V11& 	   	0.276&	0.202&	0.109&	0.051&	 4.70&	 2.67$\pm$0.43&	1.58&	 6\\
V37&  	   	0.231&	0.109&	0.073&	0.026&	 4.40&	 2.35$\pm$0.82& 5.90&	 6\\
V44&  	   	0.246&	0.171&	0.088&	0.054&	 4.93&	 3.66$\pm$0.43&	3.02&	 8\\
V45&	   	0.285&	0.303&	0.103&	0.060&	 3.94&	 2.17$\pm$1.11&	1.15&	 8\\
V62&  	   	0.128&	0.223&	0.122&	0.042&	 3.49&	 2.19$\pm$0.50&	1.11&	 7\\
V35-Field&  	0.242&	0.066&	0.062&	0.035&	 5.31&	 4.17$\pm$0.46&	2.64&	 7\\
\enddata
\label{1786ccoeff}
\end{deluxetable*}

\begin{deluxetable*}{lcccccccc}
\tablewidth{0pc}
\tabletypesize{\scriptsize}
\tablecaption{Derived Physical Properties for RRab Variables}
\tablehead{\colhead{ID}& \colhead{${\rm [Fe/H]_{J95}}$}& \colhead{$\langle M_{V}\rangle$}& \colhead{$\langle V-K\rangle$}& \colhead{$\log T_{\rm eff}^{\langle V-K\rangle}$}& \colhead{$\langle B-V\rangle$}& \colhead{$\log T_{\rm eff}^{\langle B-V\rangle}$}& \colhead{$\langle V-I\rangle$}& \colhead{$\log T_{\rm eff}^{\langle V-I\rangle}$}}
\startdata
V03& 	    	-1.625&    0.695&  1.166& 3.805&  0.332&  3.811&    0.486& 	3.812\\
V04& 	    	-1.399&    0.737&  1.073& 3.814&  0.301&  3.824&    0.446& 	3.822\\
V05& 	     	-1.346&    0.709&  1.247&  3.795&  0.380&  3.797&    0.546& 	3.796\\
V08&  	     	-1.741 &    0.764&  1.184& 3.803&  0.347&  3.807&    0.503& 	3.808\\
V10& 	     	-2.106&    0.649&  1.342& 3.787&  0.377&  3.793&     0.5431& 	3.798\\
V26& 	     	-1.120&    0.711&  1.232& 3.796&  0.388&  3.795&    0.557& 	3.793\\
V38&	     	-0.994&   0.708&  1.223&  3.796&  0.396&  3.794&    0.566& 	3.790\\
\hline
Mean&	-1.476$\pm$0.144& 0.710$\pm$0.134& 1.210$\pm$0.031& 3.800$\pm$0.001& 0.3602$\pm$0.013& 3.803$\pm$0.004& 0.521$\pm$0.017&	3.803$\pm$0.004\\
\enddata
\label{1786abphys}
\end{deluxetable*}

\section{Distance Modulus}
The absolute magnitude-metallicity relationship from \citet{catelancortes08} is used to find the absolute magnitude of the RR Lyrae stars in NGC 1786.  Since the metallicity obtained for the RRab stars, ${\rm [Fe/H]_{ZW84}}=-1.65\pm0.10$, is closer to the metallicities reported in the literature, it is used for this calculation, giving a $V$ absolute magnitude of $\langle M_{V}\rangle=0.60\pm0.22$.  Using the mean observed $V$ magnitude for the RRab stars, $\langle V\rangle=19.176\pm0.085$, the reddening value given by \citet{sharma10}, E(B-V) = $0.06$, and a standard extinction law with $A_{V}/E(B-V)=3.1$ gives a reddening-corrected distance modulus of $(m-M)_{0}=18.39\pm0.24$.  This is agrees with the distance modulus of $(m-M)_{LMC}=18.44\pm0.11$ that \citet{catelancortes08} derived for the LMC.  The location of NGC 1786 is near the center of the LMC, and the cluster distance modulus is expected to be similar to that of the LMC.  
%In order to determine if the discrepancies between these two values are a result of blending, the distance modulus is recalculated using only the six RRab stars within the cluster radius that do not have any nearby neighbors that could cause blending (V$03$, V$04$, V$05$, V$10$, V$26$, and V$27$).  The average observed $V$ magnitude for these six stars is $\langle V\rangle=19.270\pm0.075$, which changes the distance modulus to $(m-M)_{0}=18.48\pm0.23$.  This new value agrees with the distance modulus found by \citet{catelancortes08}, suggesting that the distance modulus using the full set of RRab stars was affected by blending.

\begin{deluxetable*}{lcccccc}
\tablewidth{0pc}
\tabletypesize{\scriptsize}
\tablecaption{Derived Physical Properties for RRc Variables}
\tablehead{\colhead{ID}& \colhead{${\rm [Fe/H]_{ZW84}}$}& \colhead{$\langle M_{V}\rangle$} & \colhead{$M/\msun$}& \colhead{$\log(L/\lsun)$}&\colhead{$\log T_{\rm eff}$}& \colhead{Y}}
\startdata
V07& 	     	 -1.804&     0.685& 	0.739&   1.730& 	3.864&       0.261\\
V11&  	     	 -1.819&     0.689& 	0.685&   1.734& 	3.862&       0.263\\
V37&   	     	 -2.112&     0.700& 	0.790&   1.806& 	3.853&       0.240\\
V44&   	     	 -1.820&     0.634& 	0.575&   1.742& 	3.859&       0.266\\
V45& 	     	 -1.850&      0.725& 	0.755&   1.738& 	3.863&       0.259\\
V62&   	     	 -2.139&     0.733& 	0.833&  1.827& 		3.851&       0.234\\
V35-Field&   	 -1.197&     0.672& 	0.480&   1.668& 	3.868&       0.292\\
\hline
Mean&  	-1.924$\pm$0.064&  0.694$\pm$0.014&  0.699$\pm$0.047&  1.763$\pm$0.017&  3.859$\pm$0.002&  0.259$\pm$0.007\\
\enddata
\label{1786cphys}
\end{deluxetable*}

We can also use the Type II Cepheid, V$41$, and the Anomalous Cepheid, V$01$, to check the distance modulus.  Using the period-luminosity relationships for Type II Cepheids from \citet{pritzl03} we obtain absolute magnitudes of $\langle M_{V}\rangle = -0.02\pm0.09$ and $\langle M_{B}\rangle = 0.26\pm0.14$ for V$41$.  Using the same values for reddening as we used for the RR Lyrae stars, this gives us a reddening-corrected distance modulus of $(m-M)_{0}=18.37\pm0.17$ which is consistent with what was obtained for the RR Lyrae stars.

V$01$ displays a shape that suggests that it is an Anomalous Cepheid which pulsates in the first overtone.  Using the period-amplitude relationships for first overtone ACs from \citet{pritzlarmandroff02}, we obtain absolute $V$ and $B$ magnitudes of $\langle M_{V}\rangle = -1.25\pm0.09$ and $\langle M_{B}\rangle =-1.04\pm0.12$.  This leads to a reddening-corrected distance modulus of $(m-M)_{0}=18.88\pm0.16$, which is longer than any of the other values.  If we instead assume that V$01$ is pulsating in the fundamental mode, we then obtain absolute $V$ and $B$ magnitues of $m-M_{V}=-0.45\pm0.05$ and $m-M_{B}=-0.15\pm0.06$ which yield a reddening-corrected distance modulus of $(m-M)_{0}=18.04\pm0.08$, which is shorter than any of the other values.

\section{Oosterhoff Classification}

The average periods for the RR Lyrae stars in NGC 1786 are $\langle P_{ab}\rangle=0.608$ days and $\langle P_{c}\rangle=0.336$ days.  The division between Oosterhoff-intermediate and Oo-II clusters occurs at an average RRab period of $0.62$ days; the average RRab period for NGC 1786 is just on the Oo-int side of this division.  Traditionally the average RRc period has also been used as an indicator of Oosterhoff status, and an average period of $0.335$ days would suggest an Oo-I classification for NGC 1786.  However, \citet{catelan12} recently found that average RRc period is not a strong indicator of Oosterhoff status due to the overlap in $\langle P_{c}\rangle$ values between clusters of all three Oosterhoff types; in fact an average RRc period of $0.336$ days would be consistent with any Oosterhoff classification.  
%NGC 1786's average RRc period falls on the shorter period end of the values seen for Oo-II clusters, but is still consistent with that classification.
  
We found $26$ RRab, $17$ RRc, and $8$ RRd stars within the tidal radius of NGC 1786 which gives the cluster a $N_{c+d}/N_{c+d+ab}=0.49$.  This ratio of first overtone pulsators to total number of RR Lyrae stars suggests an Oo-II classification for the cluster; though it would be similar to the higest ratio seen in Oo-int clusters.  The minimum RRab period in the cluster is $P_{ab,min}=0.49336$ days, which suggests an Oo-Int classification.

\begin{figure}
%\epsscale{0.8}
%\plotone{/data/soar/ngc1786/combined/trial/curves/v_peramp_dist.ps}
\begin{center}
\includegraphics[width=0.45\textwidth]{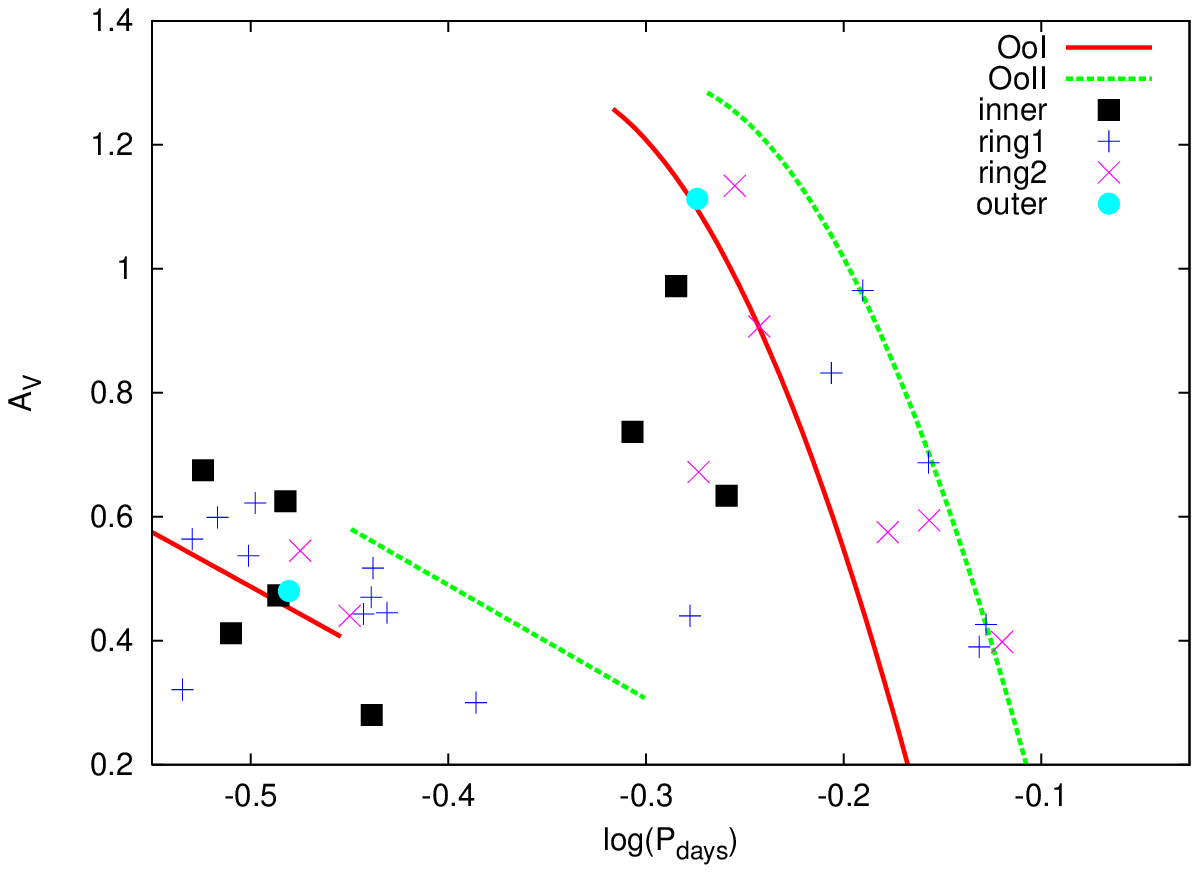}
\includegraphics[width=0.45\textwidth]{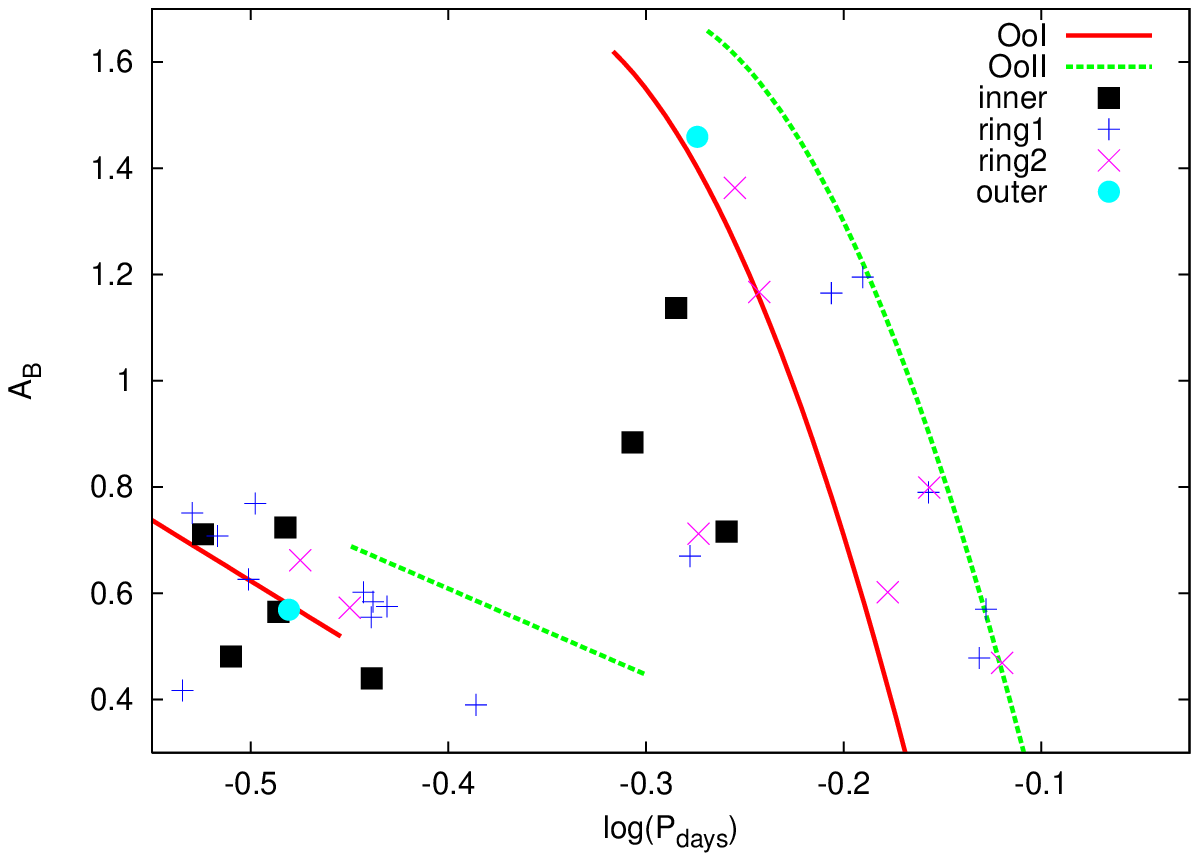}
\caption{Bailey diagram, log period vs $V$-band (top panel) or $B$-band (bottom panel) amplitude for the RR Lyrae stars in NGC 1786.  Red and green lines indicate the typical position for RR Lyrae stars in Oosterhoff I and Oosterhoff II clusters, respectively \citep{cacciari2005,zorotovic10}.  Black squares indicate stars within 21 arcseconds of the cluster center, blue plus signs indicate stars between 21 and 55 arcseconds from the center, magenta x's indicate stars farther than 55 arcseconds but still within the tidal radius of NGC 1786, while blue circles indicate stars outside of the tidal radius.}
\label{1786vperamp}
\end{center}
\end{figure}

Figure \ref{1786vperamp} shows the $V$- and $B$-band period-amplitude diagrams for NGC 1786.  Both diagrams show a great deal of scatter in the location of both the RRab and RRc stars. The scatter in the RRab stars appears to rise from two reasons.  The three RRab's in the central region of the cluster, denoted as squares, all have amplitudes that are small for their period.  An examination of their light curves shows that all three contain messy light curves that show signs of possible blending, which is not surprising considering that they are located toward the center of a crowded cluster.  Additionally, V50 has few data points near maximum light, and thus its amplitude, computed by template fitting, may be too low due to its missing maximum light in our data.  The $5$ RRab stars that were previously mentioned as being more than one magnitude too bright and showing clear evidence of blending are not included in these figures.

Looking at the RRab stars that are between 21 and 55 arcseconds from the cluster center (indicated with blue plus signs), one sees very little scatter with all the stars located near the Oo-II locus.  This region potentially provides the best representation of the RR Lyrae stars in the cluster as it is far enough from the center of the cluster that the stars should not be impacted by blending but are still likely to be cluster members.  Magenta $\times$ symbols show the RR Lyrae stars that are farther than 55 arcseconds from the cluster center but still within the tidal radius.  The RRab stars in this region show more scatter, which could be due to a greater chance of some of them being LMC field stars.  Blue circles indicate the stars that are outside of the tidal radius.

The RRc stars tend to be located closer to the Oo-I locus, which would be consistent with the average RRc periods being more similar to what is seen  in Oo-I clusters.  There does not seem to be any major shift in the position of the RRc stars with distance from the cluster center.  

\begin{figure}
\epsscale{1.0}
\plotone{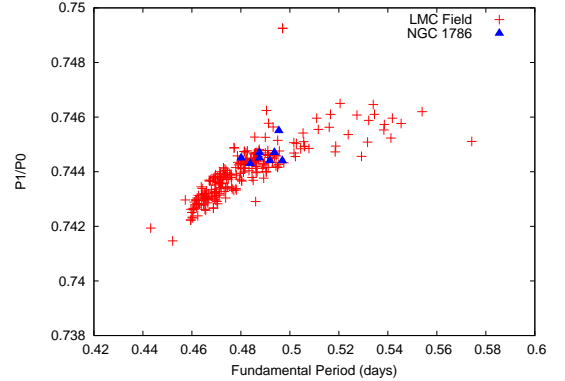}
\caption{Petersen diagram showing the ratio of the first overtone period to the fundamental-mode period vs. fundamental-mode period for the RRd stars in NGC 1786 (blue triangles).  Also plotted are the RRd stars in the LMC field (red plus symbols) from \citet{soszy03}.}
\label{1786petersen}
\end{figure}

Figure \ref{1786petersen} shows the Petersen diagram which includes the RRd stars in NGC 1786 as well as those in the LMC field.  The RRd stars in NGC 1786 fall in the same region as RRd stars in Milky Way Oo-I objects, that is, in the shorter period range ($\approx0.48-0.50$ days), while Oo-II RRd generally have longer periods ($\approx0.53-0.55$ days; see, e.g., Fig. 1 in \citet{popielski00}, and Fig. 13 in \citet{clementini04}).

\begin{figure}
\epsscale{1.0}
\plotone{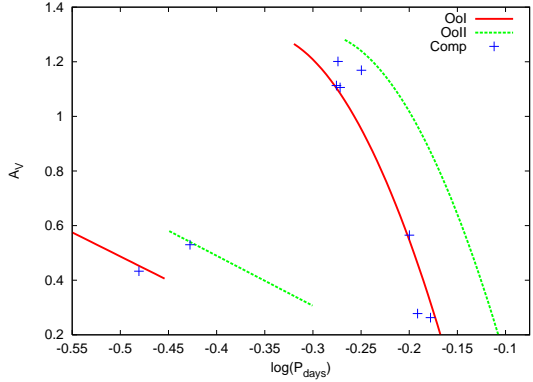}
\caption{$V$-band Bailey diagram for the RR Lyrae stars in the control field from the OGLE-III data \citet{soszy09}.  The control field is an annulus centered on NGC 1786 with the same area on the sky as the area in the tidal radius of the cluster.  The red and green lines indicate the typical position for RR Lyrae stars in Oosterhoff I and Oosterhoff II clusters, respectively \citep{cacciari2005,zorotovic10}.}
\label{compperamp}
\end{figure}

The contamination from LMC field RR Lyrae impacts the Oosterhoff classification of the cluster and without radial velocities to determine definitive cluster membership of the RR Lyrae stars it is difficult to come up with a definitive classification.  However, we can get a better handle on the Oosterhoff classification using the previously mentioned comparison field from the OGLE-III data which found $7$ RRab and $2$ RRc stars in an annulus with an area equal in size to the area within the tidal radius of NGC 1786.  Since the vast majority of the RR Lyrae stars in this comparison field were RRab stars, it is reasonable to expect that the majority of the field stars within the tidal radius are also RRab stars.  This would mean the actual value of $N_{c+d}/N_{c+d+ab}$ would be higher than the value we obtained which would be more consistent with an Oo-II classification.  Figure \ref{compperamp} shows the $V$-band period-amplitude diagram for the comparison field.  The majority of the RRab stars fall near the Oo-I locus.  Combining this with the fact that Figure \ref{1786vperamp} shows that the RRab stars near the Oo-I locus are located further from the cluster center it is reasonable to conclude that these RRab are likely to be LMC field stars and the majority of the non-blended RRab stars that are actually members of NGC 1786 are located near the Oo-II locus.

The Oosterhoff classification of NGC 1786 is more complicated than that of the other clusters examined in this study.  The first overtone dominant pulsators, the RRc and RRd stars, display properties similar to what is seen in Oo-I clusters.  However, the fundamental-mode pulsators, the RRab stars, suggest an Oo-Int classification.  The picture is further complicated by the fact that some of the RR Lyrae within the tidal radius are likely LMC field stars and should not be included when looking at the behavior of the cluster RR Lyrae stars.  The Oo-Int-like behavior of the RRab stars may be a result of this contamination.

\section{Conclusions}

We conducted a photometric study of NGC 1786 in order to identify and classify variable stars within the cluster.  A total of $65$ variable stars were found in our field of view, $53$ of which were RR Lyrae stars ($27$ RRab, $18$ RRc, and $8$ RRd).  $4$ of the variable stars ($1$ RRab, $1$ RRc, $1$ classical Cepheid, and $1$ Delta Scuti/SX Phoenicis) were outside the tidal radius of the cluster and thus are likely to be LMC field stars.  The location of NGC 1786 near the center of the LMC means that field contamination is likely within the tidal radius, however, we expect most of the variable stars found within NGC 1786's tidal radius to be cluster members.

We obtained Fourier parameters for some of the RR Lyrae stars and used these parameters to calculate physical properties of these stars.  The calculated physical properties will be compared to properties for RR Lyrae stars in other clusters in a future paper in this series.  We obtained a reddening-corrected distance modulus of $(m-M)_{0}=18.48\pm0.23$ from the six RRab stars which did not have any nearby neighbors which could cause issues with blending.

The average RRab period of NGC 1786 suggests an Oo-int classification for the cluster, though it is near the division betwee Oo-int and Oo-II, while the number ratio of first overtone dominant pulsators to total number of RR Lyrae stars suggests an Oo-II classification.  There is a great deal of scatter amongst the RRab stars on the Bailey diagram that is likely due to contamination from the LMC field stars; a comparison field suggests that the RRab stars that are cluster members likely appear more like those seen in Oo-II clusters.  The behavior of the RRc stars on the Bailey diagram and the RRd stars on the Petersen diagram both are more consistent with an Oo-I classification, while the average RRc period is consistent with either any of the three Oosterhoff types.  The behavior of the RR Lyrae on the Bailey diagram and how it compares to what is seen in other clusters will be discussed in more detail in a future paper in this series.  Physical properties for some of the RR Lyrae were obtained via Fourier decomposition of the light curves and these properties will be compared to the properties determined for the other clusters in this study in a future paper.

\acknowledgments

Support for H.A.S. and C.A.K. is provided by NSF grants AST 0607249 and AST 0707756.  M.C. is supported by Proyecto Fondecyt Regular \#1110326; BASAL Center for Astrophysics and Associated Technologies (PFB-06); the Chilean Ministry for the Economy, Development, and Tourism's Programa Iniciativa Cient\'{i}fica Milenio through grant P07-021-F, awarded to The Milky Way Millennium Nucleus; and Proyecto Anillo ACT-86.  JB is supported by the Chilean Ministry for the Economy, Development, and Tourism's Programa Iniciativa Cient\'{i}fica Milenio through grant P07-02-F, awarded to The Milky Way Millennium Nucleus and FONDECYT regular No. 1120601.

\end{document}